\documentclass[journal=jacsat,manuscript=article]{achemso}

\usepackage[version=3]{mhchem} 
\usepackage{bbold}
\usepackage{color,soul}

\newcommand{\nc}{\newcommand}
\nc{\I}{$I$}
\nc{\II}{$II$}
\nc{\III}{$III$}
\nc{\nn}{\nonumber}

\nc{\XYZ}[1]{\textcolor{black}{#1}}
\nc{\ABC }{\st}
\newcommand{\mb}[1]{\mathrm{\mathbf{#1}}}
\newcommand{\beq}{\begin{equation}}
\newcommand{\eeq}{\end{equation}}

\newcommand{\beqar}{\begin{eqnarray}}
\newcommand{\eeqar}{\end{eqnarray}}

\def\beq{\begin{equation}}
\def\eeq{\end{equation}}
\def\beqa{\begin{eqnarray}}
\def\eeqa{\end{eqnarray}}

\def\cH{{\mathcal H}}
\def\cL{{\mathcal L}}

\def\si{\sigma}

\newcommand{\tvect}[2]{%
 \ensuremath{\Bigl(\negthinspace\begin{smallmatrix}#1\\#2\end{smallmatrix}\Bigr)}}

\author{Oliver L. A. Monti}
\email{monti@arizona.edu}
\affiliation[UoA]
{Department of Chemistry and Biochemistry, University of Arizona, 1306 E. University Blvd., Tucson, Arizona 85721, United States \\Department of Physics, University of Arizona, 1118 E. Fourth Street , Tucson, Arizona 85721, United States}

\author{Yonatan Dubi}
\email{jdubi@bgu.ac.il}
\affiliation[BGU]
{Department of Chemistry, Ben Gurion University of the Negev, 1 Ben-Gurion Ave, Beer Sheva, 8410501, Israel}

\title{Surface magnetic stabilization and the photo-emission  chiral-induced spin-selectivity effect}

\abbreviations{CISS}
\keywords{Chiral Induced Spin Selectivity, Spinterface}

\begin{document}

\begin{tocentry}
\includegraphics[width=0.8\linewidth]{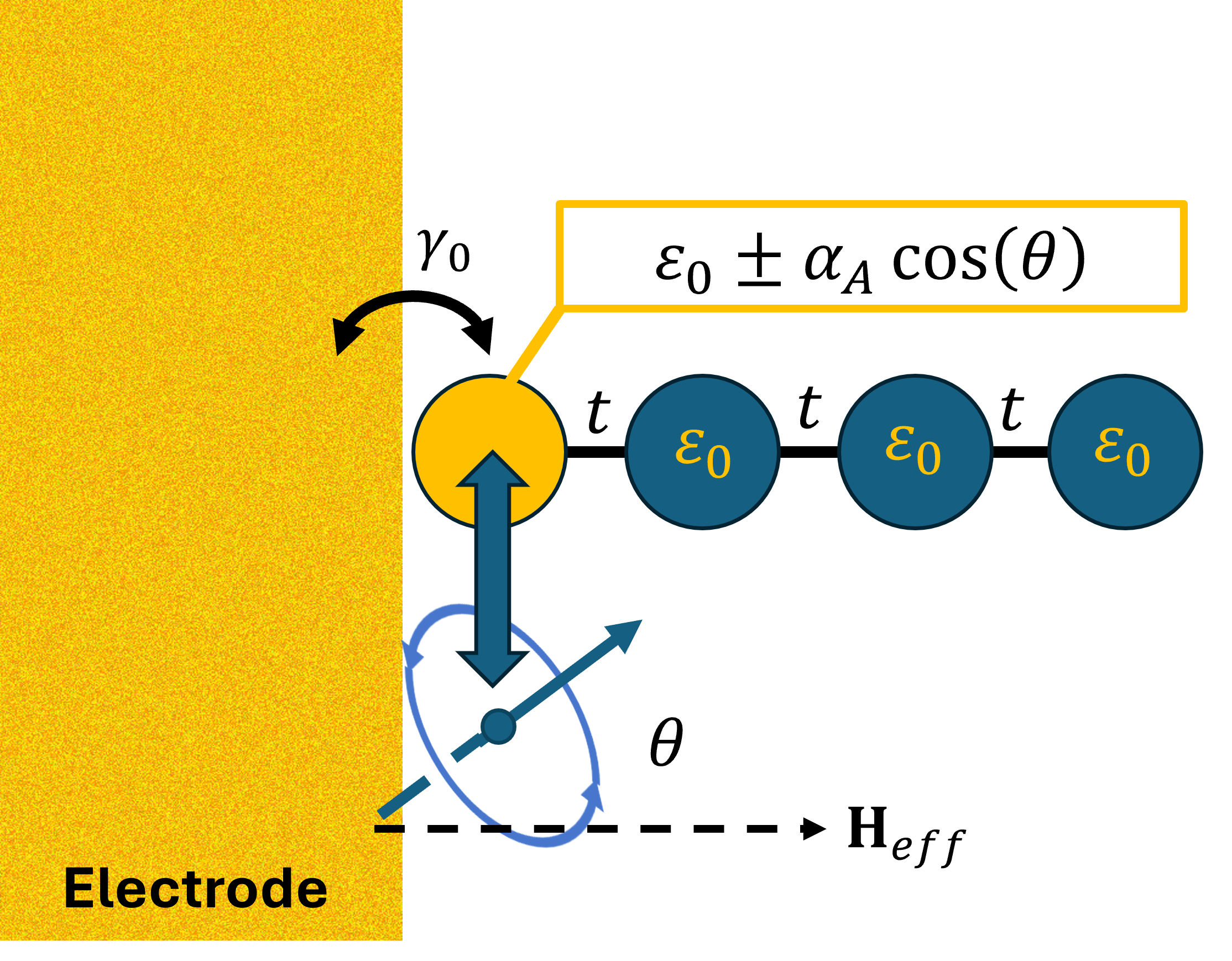}

\end{tocentry}

\begin{abstract}
The spinterface mechanism was suggested as a possible origin for the chirality induced spin-selectivity (CISS) effect, and was used to explain and reproduce, with remarkable accuracy, experimental data from transport experiments showing the CISS effect. Here, we apply the spinterface mechanism to explain the appearance of magnetization at the interface between non-magnetic metals and chiral molecules, through the stabilization of other-wise fluctuating magnetic moments. We show that the stabilization of surface magnetic moments occurs for a wide range of realistic parameters and is robust against dephasing. Importantly, we show that the direction of the surface magnetic moments is determined by the chiral axis of the chiral molecules. Armed with the concept of stable surface magnetic moments, we then formulated a theory for the photoemission CISS effect. The theory, based on spin-dependent scattering, leads to direct predictions regarding the relation between the photoemission CISS effect, the chiral axis direction, the spinterface "size", and the tilt angle of the detector with respect to the surface. These predictions are within reach of current experimental capabilities, and may shed new light on the origin of the CISS effect.   
\end{abstract}

\section{Introduction}

The chirality-induced spin-selectivity (CISS) effect encompasses a range of phenomena characterized by an apparent splitting of the otherwise spin degenerate electronic states of a solid in the presence of chiral molecules and in response  to external stimuli.\cite{aiello2022chirality, xu2023chiral,aragones2022magnetoresistive,Bloom2024chiral}. There are two central manifestations most commonly associated with CISS: 1) Transport experiments \cite{Naaman12,Naaman15,naaman2019chiral,naaman2020chiral}, where current flows through a junction composed of a normal metal electrode decorated by one or many chiral molecules and contacted by a ferromagnetic electrode. The current-voltage characteristics vary depending on whether the ferromagnet is magnetized parallel or anti-parallel to the direction of current flow, which typically (or presumably) aligns with the molecular chiral axis. 2) Photoemission, which constitutes the original observation of CISS \cite{mollers2022chirality,kettner2018chirality,mollers2022spin,Gohler11,abendroth2019spin,badala2022vectorial,nino2014enantiospecific}. In photoemission experiments, electrons are photoexcited to leave a non-magnetic metal surface decorated with chiral molecules. Remarkably, the photoemitted electrons emerging from the surface show some degree of spin polarization, which depends on the handedness of the chiral molecules. Typically (but not exclusively \cite{yang2023}), CISS is observed with helical molecules such as polypeptides, oligonucleotides (i.e. short, single- or double-stranded DNA or RNA) or small helical organic molecules such as helicenes. In all these, the helical axis is also the chiral axis, imagined to be normal to surface.  

Though both transport and photoemission through chiral molecules are usually viewed as different manifestations of the same underlying physical mechanism of CISS, there are fundamental differences between the two, which may at least in principle require different explanations: Transport studies probe non-equilibrium phenomena and take place in the presence of a ferromagnetic electrode, such that time-reversal symmetry is inherently broken in the system under study. The question of how CISS works then amounts to an investigation on how chiral molecules influence the spin-dependent transmission in some nanoscale device. Additionally, in transport experiments the measurement implicitly defines a preferred direction, namely the transport axis, which usually coincides with the chiral (helical) axis. 

In contrast, the situation in photoemission experiments may be more subtle: CISS and the associated observed spin polarization may stem from spin-dependent transmission through a chiral molecule, similar to transport experiments, or  photoemission may probe a preexisting interfacial magnetization created by the chiral molecules. The latter would require breaking of time-reversal symmetry in equilibrium and in the absence of ferromagnetic electrodes. Support for this surprising interpretation comes from a variety of experiments that report changes in surface magnetization upon adsorption of chiral molecules \cite{ben2017magnetization,koplovitz2019single,metzger2020electron,tassinari2018chirality,abendroth2019spin, abendroth2017analyzing,Bloom2024chiral,mishra2024inducing,theiler2023detection,Nguyen2024}. In addition, there is no predefined symmetry axis in photoemission, since electron take-off angles can vary greatly, from surface normal to close to along the surface plane, and the chiral axis need not be aligned with the surface normal.

From a theoretical perspective, the CISS effect and its origin(s) are considered an open question \cite{Evers20,naaman2020chiral,Bloom2024chiral,liu2023spin,tirion2024mechanism}. While only a few theoretical studies address the CISS transport phenomenon and transmission of electrons through chiral molecules during photoemission \cite{medina2012chiral,eremko2013spin, Ghazaryan20,varela2013inelastic}, there are even fewer papers that tackle the more striking proposed generation of magnetization of metallic surfaces by chiral molecules, i.e. the equilibrium breaking of time-reversal symmetry in the absence of external magnetic fields \cite{fransson2021charge,shiranzaei2023emergent,fransson2023vibrationally,fransson2022charge}. Here, we fill this gap by developing a dynamical version of the recently-proposed spinterface mechanism for the CISS effect  \cite{alwan2021spinterface}. We show that even in the absence of transport, adsorption of chiral molecules creates a stable equilibrium interface magnetization (i.e. a "spinterface"). We then deploy this formalism to shed light in a simple and physically transparent yet quantitatively satisfactory model for how the CISS effect arises in photoemission. This model also fully treats the in principle arbitrary chiral axis and predicts CISS magnitudes that agree well with the best experimental data available. Remarkably, it also predicts a length-dependence of the magnitude of CISS, long considered a hallmark of CISS.

Unlike many proposed models for CISS that are unable to explain the magnitude of the observed spin selectivity, the spinterface mechanism, suggested some time ago \cite{alwan2021spinterface}, is able to quantitatively explain a range of CISS observations. It was successfully used to  describe raw data of the CISS effect in transport experiments involving biological chiral molecules \cite{dubi2022spinterface} and small molecules in single-molecule junctions \cite{yang2023}, intercalated chiral layers \cite{alwan2023temperature}, and superhelical PANI microfibers \cite{alwan2024role}. It also sheds light on the temperature-dependence of the CISS effect \cite{alwan2023temperature}, and provides theoretical bounds for the expected CISS polarization  \cite{alwan2024role}. Here, we extend this model to explain the emergence of surface magnetization in otherwise nonmagnetic surfaces decorated with chiral molecules, and offer a simple model to understand CISS in photoemission by investigating the dynamical stabilization of interface magnetic moments at such interfaces.

The main insight offered here is that when chiral molecules are placed on a surface, electrons spontaneously move across the interface due to the molecule-metal contact and the overlap between the surface and the molecular wavefunctions. The resulting displacement currents in or out of {\it chiral} molecules stabilize the otherwise randomly oriented magnetic moment at the surface, analogous to the more familiar situation in transport experiments. Consequently, we propose to interpret the emergence of spin polarization in photoemission from interfaces with chiral molecules - treated here explicitly - to arise from the creation of a spinterface magnetization that persists once equilibrium is established after adsorption of chiral molecules.

\section{The dynamical spinterface model}

Since the spinterface mechanism was described elsewhere in detail \cite{alwan2021spinterface,yang2023,dubi2022spinterface}, we only briefly outline the physical basis here. The central idea is that the surface of a paramagnetic metal (typically gold) can support magnetization, yet without chiral molecules or some mechanism that breaks time-reversal symmetry the surface magnetization fluctuates spatially and/or temporally and thus averages out to zero. The spinterface mechanism, so far only discussed for transport scenarios, is the process of stabilization of this surface magnetization due to spin-torque (or spin-exchange) cross-interface interactions between the surface moments and the electron spins in the molecule. The directional symmetry is initially broken by the effective solenoid field in the chiral molecule, generated as electrons flow through the molecule. A cascade of energy scales (from the small solenoid field, through the intermediate spin-torque, up to the metal electrode SO interactions) amplifies the very small solenoid effect and allows for CISS polarization to reach very high values.  \cite{alwan2024role} 

\begin{figure}
\begin{center}
\includegraphics[width=12.6cm]{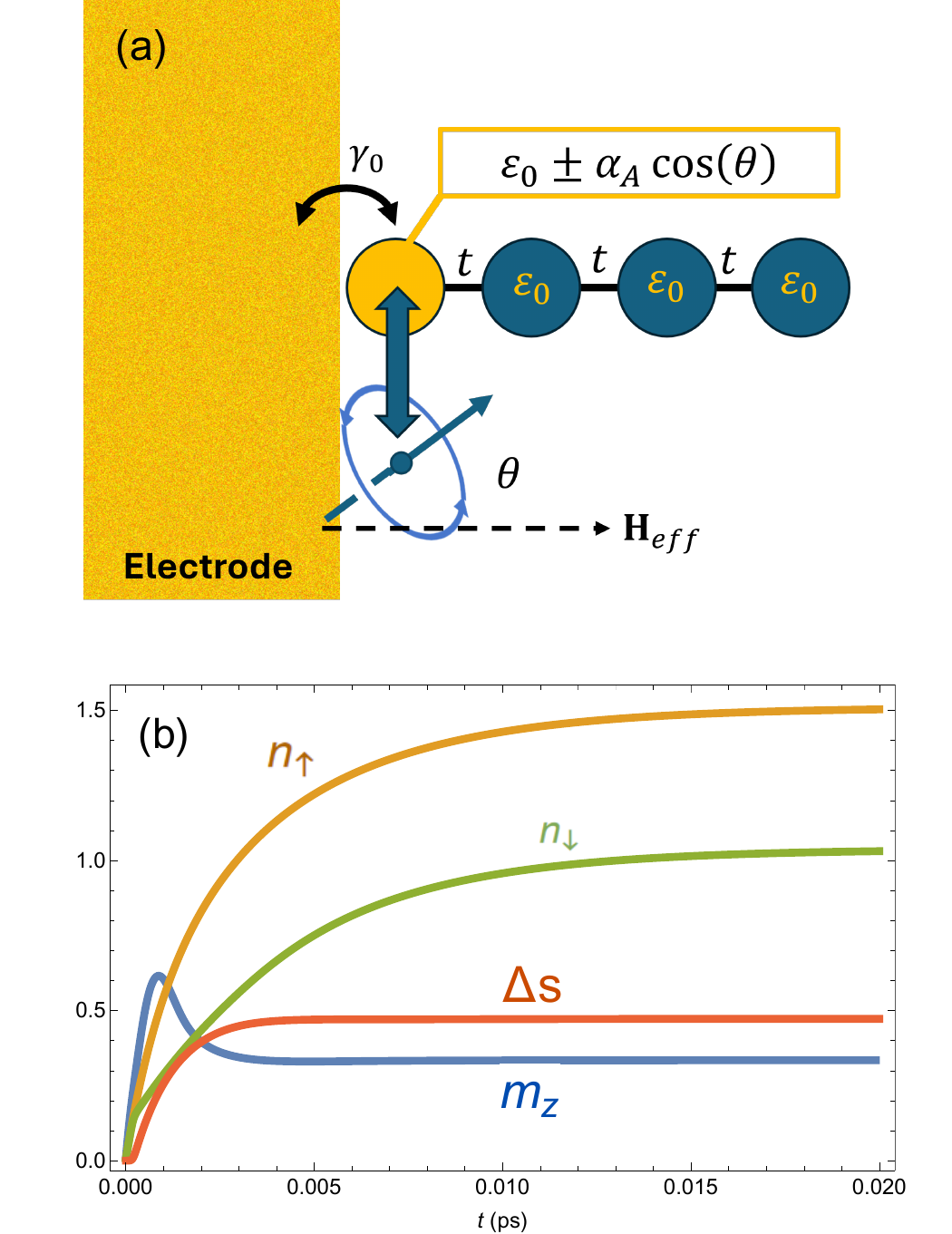}
\caption{(a) Schematic presentation of the system under consideration, composed of a molecule (circles) placed on an electrode surface. The electrons flow from the electrode to the molecule, and at the contact site (orange circle) undergo a Zeeman splitting due to the interaction with a localized moment (blue arrow). The localized moment feels an effective field (dashed arrow) due to the solenoid field in the chiral molecule and a spin-imbalance formed in the molecule. These two coupled processes lead to the stabilization of the surface moment and to generation of finite spin-imbalance and hence magnetic moment in the molecule. (b) Density and magnetization dynamics: $m_z$ (blue), $n_\uparrow$ (orange), $n_\downarrow$ (green) and $\Delta s =n_\uparrow-n_\downarrow$ as a a function of time. As the molecule is attached to the electrode at time $t=0$, both spin species start moving into the molecular empty states, generating a displacement current, which tilts the magnetic moment towards the $z$ direction (the molecular chiral axis). The feedback between $m_z$ and $\Delta s$ (Eqs.~ \ref{E_isigma} - \ref{Hef}) leads to a full alignment of $m_z$ with the molecular chiral axis \XYZ{(finite value for $m_z$)} and to a substantial difference in spin-occupation in the molecule. }
\label{Fig_schematic}
\end{center}
\end{figure}

Here we study the transient effect of placing chiral molecules on a gold surface \cite{fransson2022charge}, newly taking into account the displacement current of electrons through the interface and the interaction of the electrons in the chiral molecule with a localized surface magnetic moment (a schematic description is shown in Fig.~\ref{Fig_schematic}). When molecules are placed on the surface, electrons will move from the metal to the molecule \cite{monti2012understanding} (or vice-versa, depending on the position of the molecular HOMO or LUMO compared to the metal Fermi level, but the formalism and the results are qualitatively identical for both cases). This displacement current, as it flows through the chiral molecule, will generate a small solenoid field which breaks the directional symmetry. Through the spinterface mechanism, this field is enhanced by the spin-torque interaction, leading to a stabilization of the surface moments, and to the formation of spin-imbalance and hence a magnetic moment in the molecule. 

To show this, one needs a theoretical description that couples electron motion from the metal into the molecule with the dynamics of surface moments. We consider the system schematically described in Fig.~\ref{Fig_schematic}, and describe the electronic motion using Lindblad equations \cite{lindblad1976generators, pearle2012simple, breuer2002theory, manzano2020short} and the magnetic moment dynamics with a Landau–Lifshitz–Gilbert equation (LLG) equation \cite{landau1992theory, gilbert2004phenomenological, lakshmanan2011fascinating,stiles2006spin}. 

The Lindblad equation describes the dynamics of the electrons in the molecule interacting with the reservoir of electrons in the metal contact, and is of the form $\dot{\rho}(t)=-\frac{i}{\hbar}[\cH,\rho]+\cL[\rho]$, where $\rho$ is the electronic density matrix, $\cH$ is the molecular Hamiltonian of an assumed linear chain-like molecule, and $ \cL[\rho] =\sum_k \left(V_k\rho V^{\dagger}_k-\frac{1}{2} \left\{ V^{\dagger}_k V_k,\rho \right\}\right)~$
is the so-called Lindbladian, where $V_k$ are a set of operators describing the action of the reservoir on the electronic system. Here $\hbar$ is the Planck constant, and $[,]$ and $\{,\}$ are the commutation and anti-commutation relations, respectively. The Hamiltonian is a simple tight-binding Hamiltonian $\cH=\sum_{i\si} \varepsilon_{i\si} c^\dagger_{i\si} c_{i\si}-t\sum_{\langle ij \rangle ,\si}  c^\dagger_{i\si} c_{j\si}$, where $c^\dagger_{i\si} (c_{i\si})$ creates (annihilates) an electron at molecular site $i$ with spin $\si$, $t$ is the hopping matrix element, and we consider only a spin-conserving nearest-neighbor hopping.  

The Lindblad terms encode hopping of electrons from the metallic electrode into the contact site in the molecule, which, for simplicity, is taken to be the left-most site in the chain with index $i=1$. A simple model for such a term \cite{dubi2009thermoelectric,dubi2011colloquium,ajisaka2015molecular,sarkar2020environment} is described by hopping of an electron from a Fermi reservoir onto a single particle state $\psi_k$ with eigen-energy $\varepsilon_{k\si}$, of the form  $
V_{k,in}= \gamma_0^{1/2} \sqrt{|\psi_k(1)|^2 f_{FD}(\varepsilon_{k,\si}) } c^\dagger_{k\si} 
$ and the reverse process $
V_{k,out}= \gamma_0^{1/2} \sqrt{ |\psi_k(1)|^2 (1-f_{FD}(\varepsilon_{k,\si}))} c_{k\si} ,~$
where $f_{FD}(\varepsilon)$ is the Fermi function at temperature $T$ and setting the chemical potential to be zero,  and $k$ runs over all molecular single particle states. $\gamma_0$ is the rate at which electrons cross the metal-molecule interface (typically ($1-10$fs)$^{-1}$), $\psi_k(1)=\langle 1 |k\rangle$ is the overlap of state $k$ with the real-space orbital at site $1$ (the site connecting the molecule to the electrode) and the term $|\psi_k(1)|^2$ makes sure that only states with weight at the contact point can be populated by the electrode. This description is thermodynamically consistent \cite{dubi2009thermoelectric,dubi2011colloquium}.

The only place where the different spin species play a role in the above description is in the on-site energies $\varepsilon_{i\si}$. The spinterface model implies that the surface magnetic moments interact with the electrons in the molecule. If we assume that an interface magnetic moment is described by a magnetization $\mb{M}=M\{m_x,m_y,m_z\}$, where $M$ is the magnitude of the magnetic moment $~\mu_0 \hbar$, then one can approximate the single-electron energies as \cite{alwan2021spinterface}
\begin{equation}
 \varepsilon_{i\si}=\varepsilon_{i0} + \delta_{i,1}\sigma \alpha_A m_z ~.\label{E_isigma}
 \end{equation}
  Here $\varepsilon_{i0}$ is the bare level energy, $\alpha_A$ is the magnitude of the spin-orbit interaction energy in the electrode, typically $\sim 0.1-1$eV e.g. in Au, and $m_z=\cos(\theta)$ is the projection of the magnetic moment on the $z$-direction. We start by assuming that the magnetic moment averages out to zero in the $x-y$ plane \cite{alwan2021spinterface}, and will relax this restriction in the following sections. The form of $\varepsilon_{i\si}$ implies that in the description of the system, only site $\#1$ belongs to the electrode and hence feels the metal SO interaction  \cite{alwan2021spinterface}, and the bare level undergoes a Zeeman splitting due to its interaction with the surface magnetic moment. 

Several things are noteworthy at this point. First, we note that the values of the SO interaction assumed for gold are the atomic values, and these are considerably smaller for surface states \cite{nicolay2001spin}. However, it has already been demonstrated that the spinterface mechanism for the CISS effect continues to work even for much smaller values of the SO interaction  \cite{dubi2022spinterface}. Second, it is to be understood that the first site in the "chain" represents the last gold atom connected to the molecule. Thus one could argue that it should have different local energy and coupling. However, this will only make a qualitative change in the results, as verified numerically, and add additional parameters which need to be specified. Thus, for the sake of simplicity, the parameters of the chain are taken to be uniform. \XYZ{Third, we wish to clarify that although we refer to the interaction between the surface magnetic moment and the electron spins as SO interaction, it very well may be that this is instead spin-exchange interaction, because the surface magnetization can arise from either the orbital angular momentum of localized surface states or from the spins of electrons occupying these localized surface states. The mean-field  Hamiltonian is the same in both cases and leads therefore to the same dynamics.  }

The dynamics of the surface magnetic moment are determined by using the Lifshitz-Landau-Gilbert (LLG) equation \cite{landau1992theory, gilbert2004phenomenological, lakshmanan2011fascinating,stiles2006spin}, 
\beq
\frac{d \mathrm{\mathbf{m}}}{dt}=-\gamma \mathrm{\mathbf{m}} \times \mb{H}_{eff}-\lambda \mb{m} \times (\mb{m} \times \mb{H}_{eff})
\eeq
where $\mb{m}(t)$ is the normalized surface magnetic moment, $\gamma$ is the gyromagnetic ratio and $\lambda$ is the Gilbert damping factor. The important component of the LLG equation, coupling the dynamics of the surface moment to the electron dynamics, is the effective field $\mb{H}_{eff}$, which is given by (multiplied by $M$) as\cite{alwan2021spinterface} \beq M\mb{H}_{eff}=\left(\alpha_0 J +\alpha_1 (n_\uparrow-n_\downarrow)\right) \hat{n}\label{Hef}\eeq  in the direction $\hat{n}$ defined by the chiral axis of the molecule, and $J$ is the total current inside the molecule, calculated from the density matrix \cite{zerah2018universal}. The first term has its origin in the molecular chirality, and corresponds to the generation of a "solenoid field" due to $J$, the motion of electrons in the chiral molecule. It is typically very small, and effectively serves only for breaking the directional symmetry. Put differently, this is the term which couples between the chirality (left or right) to the majority spin species (parallel or anti-parallel to the z-axis, or to any other axis of choice). The second term arises from the spin-exchange or spin-torque interaction across the interface and is the term which allows for the small solenoid field to stabilize the orientation of the surface magnetization. The energy scale cascade $\alpha_0 \langle J \rangle  \sim 10^{-5}\mathrm{eV} \longrightarrow \alpha_1 \sim 10^{-2}\mathrm{eV} \longrightarrow \alpha_A\sim 1\mathrm{eV}$ leads to the possibility of a macroscopically observable CISS effect. 

In all results shown in what follows, we set the following numerical parameters: the molecular length is $L=3$, $\varepsilon_0=-0.2$ and $t=1.6$ (all energy scales are in eV, and we set the chemical potential of the electrode as the zero). We take room temperature, set $\gamma_0=1$ fs$^{-1}$, and set the Gilbert damping to be $\lambda=1 \gamma$. This latter choice only affects the results qualitatively. For the CISS parameters we take $\alpha_A=0.85$eV, $\alpha_0=10^{-7}$T/nA. Unless stated otherwise, these parameters are fixed throughout the calculations. We note that these parameters are similar to those chosen in the original implementation of the spinterface model \cite{alwan2021spinterface}.

\section{Results: CISS dynamics}

The basic dynamics of the system are shown in Fig.~\ref{Fig_schematic}(b), where $m_z$ (blue line), $n_\uparrow$ (orange line), $n_\downarrow$ (green line) and $\Delta s =n_\uparrow-n_\downarrow$ (red line) are plotted as a a function of time, \XYZ{taking $\alpha_1=8\times 10^{-2}$eV}. Here we take the molecular axis to be in the $z$-direction, i.e. $\hat{n}=z$. At time $t=0$, the molecule is coupled to the metallic surface, where the surface magnetization is pointing in the $x$ direction. Note that it can point in any arbitrary direction without affecting the overall results. As electrons start to flow from the electrode to the molecule, a displacement current is generated. The interplay between the solenoid field, which breaks the symmetry and sets the direction of the magnetization, the magnetization $m_z$ and the spin-difference $\Delta s$ through the spin-torque and the electrode SO coupling, (Eqs.~ \ref{E_isigma} - \ref{Hef}) lead to mutual enhancement of $m_z$ and $\Delta s $. The dynamical feedback between $m_z$ and $\Delta s$ (Eqs.~ \ref{E_isigma} - \ref{Hef}) leads to an alignment of $m_z$ with the molecular chiral axis. \XYZ{ After sufficient time, it reaches a new equilibrium state, characterized by a stable magnetization at the interface (finite non-zero value for $m_z$, in the example of Fig.~\ref{Fig_schematic}(b), $m_z\sim0.3$, corresponding to a magnetization angle of $\theta\sim 0.1 \pi$}) and a finite spin density in the molecule. This shows unambiguously that the presence of a chiral molecule at an interface results in stabilization of the direction of the magnetic moment. 

\XYZ{We point out that, as expected, the magnitude of the solenoid field (i.e. $\alpha_0$) does not affect the dynamics, since the solenoid field only breaks the spin symmetry and determines the majority spin species. Consequently, when the calculations are performed with a negative value of $\alpha_0$, which represents flipping the molecular chirality, the dynamics are exactly the same but with reversed roles for the up and down spins. This is a central hallmark of the CISS effect. }

\XYZ{The fast dynamics of a few fs arise from two reasons: i) The electron dynamics  are determined by the coupling between the electronic states of the molecule and the metal, which for many cases defines a time-scale of 1-10fs \cite{nitzan2001electron,zhu2004electronic}. ii) The magnetization dynamics are defined by the interaction energy scale $\alpha_1\sim 50$meV, which, if translated to a magnetic field, correspond to a magnetic field pulse of $\sim100$T, thus leading to the fast dynamics.}

\begin{figure}[h]
\begin{center}
\includegraphics[width=12.6cm]{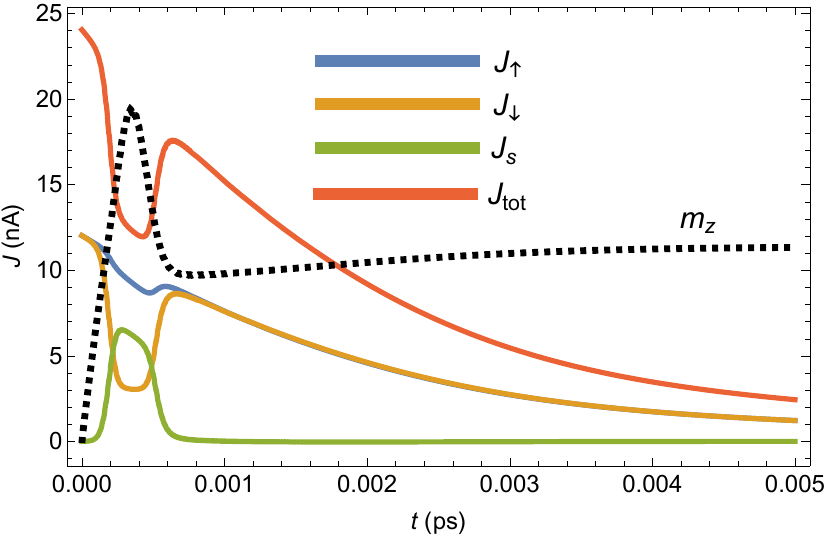}
\caption{\XYZ{Time-dependent currents $J_{\uparrow,\downarrow}$ of up-electrons and down-electrons, respectively, the spin-current $J_s=J_\uparrow-J_\downarrow$, the total current $J_s=J_\uparrow+J_\downarrow$ and the surface magnetization $m_z$ (arbitrary units, dashed black line), demonstrating the dynamical interplay between the displacement currents and the surface magnetization. Parameters are the same as in Fig.~\ref{Fig_schematic}(b).}}
\label{Fig_DisplacementCurrents}
\end{center}
\end{figure}

\XYZ{In order to see more clearly the dynamical relation between the spin-separated displacement currents and the stabilization of the surface magnetization, in Fig.~\ref{Fig_DisplacementCurrents} we show the time-dependent currents $J_{\uparrow,\downarrow}$ of up-electrons and down-electrons respectively, the spin-current $J_s=J_\uparrow-J_\downarrow$ and the total current $J_s=J_\uparrow+J_\downarrow$ (see color coding in the figure), all calculated from the time-derivative of the respective densities. For comparison, the magnetization $m_z$ is also plotted. One can clearly see that the interplay between the currents and the magnetization, all of which exhibit similar dynamics with a dynamical "blip" around $\sim5$fs, where the moment stabilizes. }

\begin{figure}
\begin{center}
\includegraphics[width=12.6cm]{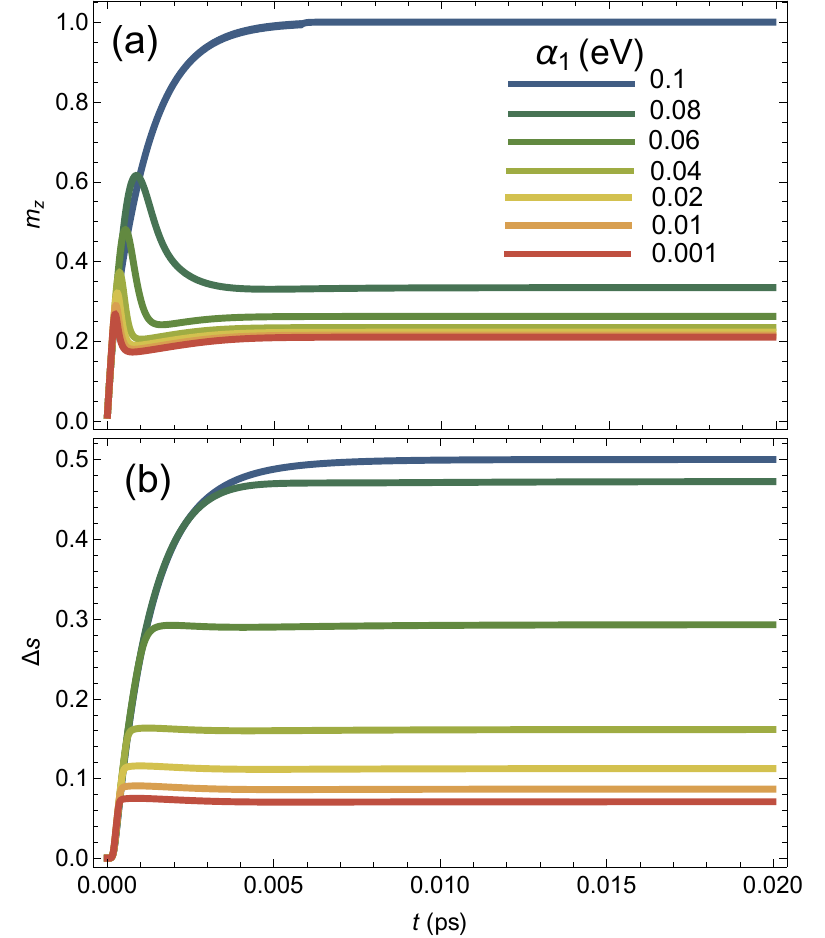}
\caption{(a) Surface moment, $m_z$, as a function of time for various values of $\alpha_1$, ranging from $\alpha_1=10^{-3}$eV to $10^{-1}$eV (see legend). Above some critical value of $\alpha_1$, the surface magnetization stabilizes at $m_z=1$. (b) Spin-imbalance $\Delta s$ as a function of time, for same values of $\alpha_1$. }
\label{Fig_2}
\end{center}
\end{figure}

In Fig.~\ref{Fig_2} the magnetization and density dynamics are shown for different values  of $\alpha_1$, ranging from $\alpha_1=10^{-3}$eV to $10^{-1}$eV (see legend). As can be seen, only above some critical value of $\alpha_1$ can the magnetization stabilize at $m_z=1$, the fully developed CISS effect. The reason is that if $\alpha_1$ is not large enough, the spin-imbalance developed in the molecule is not large enough to stabilize the surface moment, and the density saturates in the molecule before the surface moment is fully directed to the direction of the chiral axis. A similar effect would be observed if $\alpha_A$ is reduced. Nevertheless, significant spin imbalance can be expected even for smaller values of $\alpha_1$ and $\alpha_A$.

In the calculation above, while the dynamics of the surface magnetisation are classical, the electron dynamics inside the molecule are fully coherent. This is an approximation which may not be valid in, e.g., long molecules at room temperature, where soft vibrations lead to dephasing and to essentially classical dynamics of the electrons. Within the Lindblad equation, a crossover between coherent and incoherent dynamics can easily be studied, by encoding all dephasing processes into local dephasing terms of the form $V_{deph,i}=\sqrt{\gamma_{deph}} c^\dagger_{i\si}c_{i\si}$ \cite{breuer2002theory}, where $\gamma_{deph}$ is the dephasing rate. Increasing dephasing takes the system from purely quantum coherent dynamics to essentially classical (secular) dynamics \cite{zerah2018universal,dziarmaga2012non}. 

\begin{figure}
\begin{center}
\includegraphics[width=16.6cm]{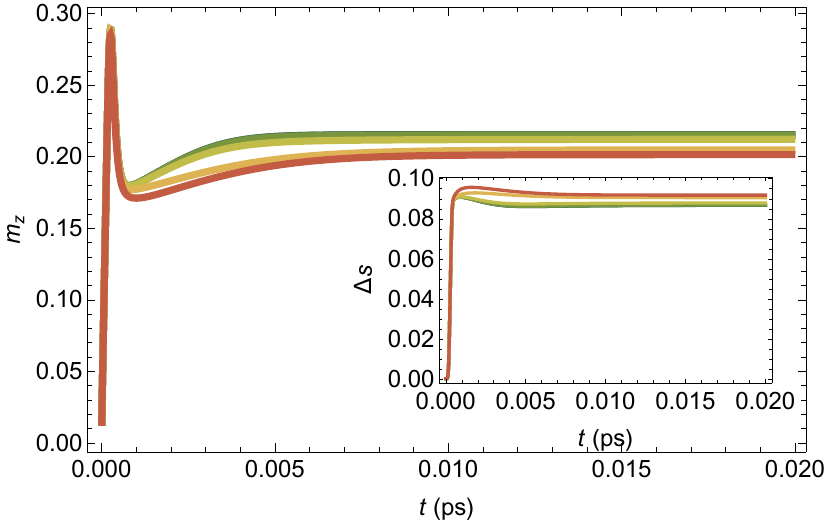}
\caption{Magnetization $m_z$ and spin imbalance $\Delta s$ (inset), as a function of time, for different values of the dephasing rates, corresponding to dephasing timescales of $1$ps,$0.1$ps,$100$fs,$10$fs,$1$fs and $0.1$fs (from green to red lines, respectively), showing indifference to dephasing.  }
\label{Fig_deph}
\end{center}
\end{figure}

In Fig.~\ref{Fig_deph} the magnetization $m_z$ and the spin imbalance $\Delta s$ (inset to Fig.~\ref{Fig_deph}) are plotted as a function of time for different values of the dephasing rate, $\gamma_{deph}=0-10^4$ ps$^{-1}$ \XYZ{(see color-coding in the figure caption)}. As can be seen, dephasing has virtually no effect on the dynamics of the magnetization and spin imbalance. Put simply, the dynamical spinterface mechanism described above works equally well for either coherent or incoherent electron dynamics. Note, however, that the dynamics of the surface magnetization are always classical, and contain inherent dissipation through the Gilbert damping term. 

\section{Results: Molecular angle dependence}

In realistic systems, the chiral axis of molecules attached to the surface does not necessarily point in the $z$-direction, i.e. normal to the surface. Whether in a self-assembled monolayer or in isolated molecules, it is more likely that there will be some angle $\theta$ between the $z$-direction and the molecular chiral axis. In fact, in chiral molecules, even if the main backbone axis of the molecule is pointing directly along the $z$-direction, it is not at all guaranteed that the "effective solenoid" direction will be in that same direction. This is because the latter is generated by a helical current, which may not be directed along the main molecular axis, especially if the chirality of the molecule is generated by some chiral side-group \cite{yang2023} as opposed to helical molecules such as helicenes \cite{kiran2016helicenes}. 

\begin{figure}
\begin{center}
\includegraphics[width=16.6cm]{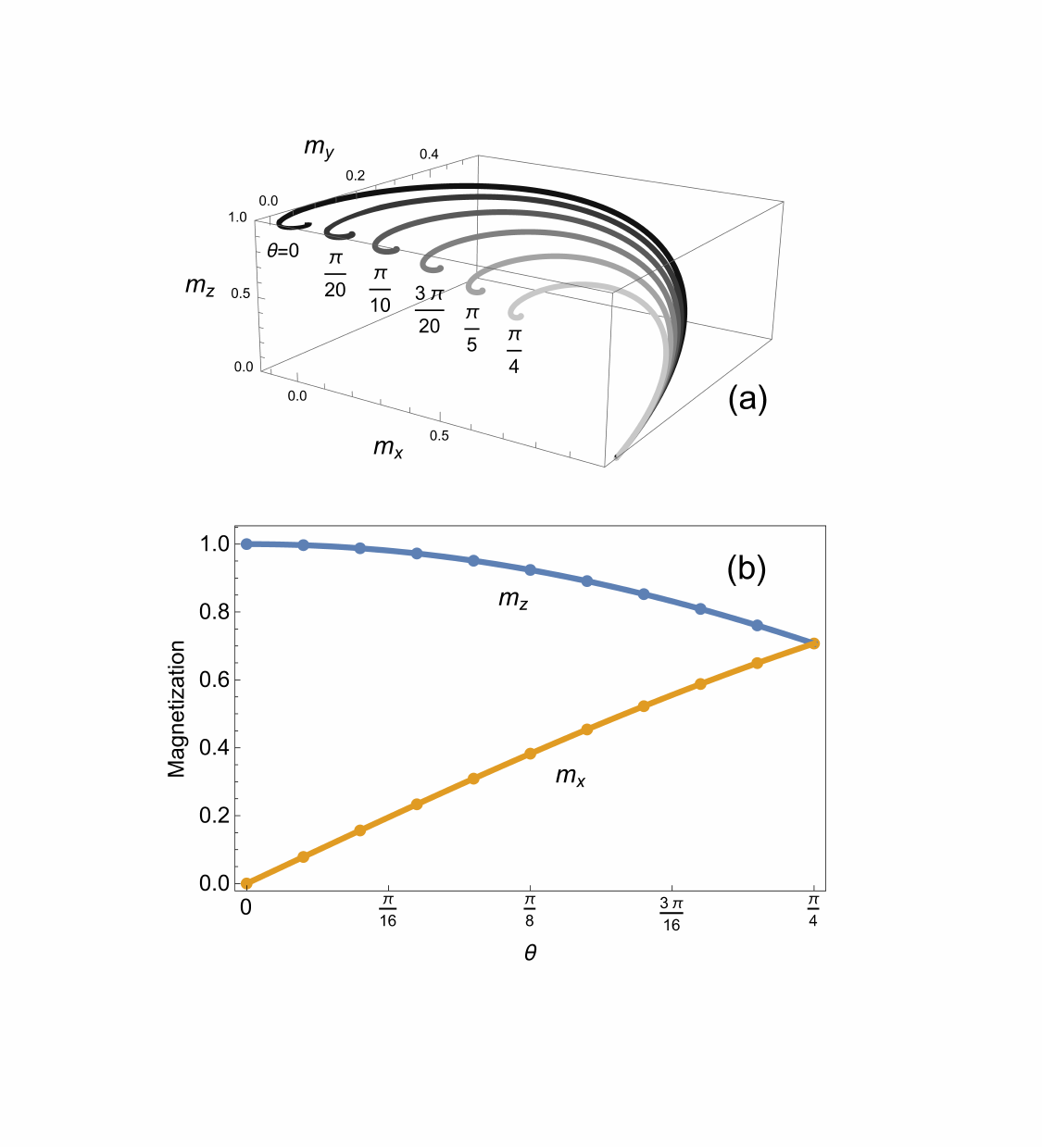}
\caption{(a) Parametric plot of the surface magnetization $\{m_x(t),m_y(t),m_z(t)\}$ (with  $\mb{m}(t=0)=\{1,0,0\}$ ), for different values of $\theta$. (b) The long-time limit values of $m_z$ (blue) and $m_x$ as a function of $\theta$. Circles are calculated from the dynamics, and the solid lines are $\cos(\theta)$ (blue) and $\sin(\theta)$ (orange).   }
\label{Fig_3}
\end{center}
\end{figure}

  In Fig.~\ref{Fig_3}(a) a parametric plot of the surface magnetization $\mb{m}(t)=\{ m_x(t),m_y(t),m_z(t) \}$ is plotted for different values of $\theta$, i.e. when the chiral molecular axis is pointing in a direction $\hat{n}=\{\sin (\theta),0,\cos (\theta)\}$. Without loss of generality, we set the initial conditions to $m_x=1,m_y=m_z=0$. As seen, the LLG dynamics lead to stabilization of the surface moments, but the surface magnetization does not saturate to $m_z=1$. In Fig.~\ref{Fig_3}(b) we plot the long-time limit $m_z$ (blue points) and $m_x$ (orange points). The solid lines are  $\cos(\theta)$ (blue) and $\sin(\theta)$ (orange), demonstrating that, the surface moments align with the molecular chiral axis. \textcolor{black}{This not only broadens the class of molecules that should be considered suitable for CISS, but also offers an opportunity to control the principal direction of excess spin, useful e.g. to control the direction of spin-torque in spintronic devices.}

\section{Application to Photoemission CISS experiments}

The key result of the dynamical CISS calculations can be summarized as follows: When chiral molecules are placed on a metallic surface, a surface magnetization is stabilized in the direction of the molecular chiral axis, provided spin-torque and SOC are strong enough. With this conclusion, we can now construct a simple yet illustrative and to some extent quantitative theory for the photoemission CISS effect. 

The CISS photoemission experiments are typically performed with rather low illumination intensity \cite{mollers2022chirality,nino2014enantiospecific,badala2022vectorial,kettner2018chirality,mollers2022spin2}, implying that the total number of emitted electrons per molecule and unit time is much smaller than current experienced by molecules in transport CISS experiments. This points to the possibility that the origin of the CISS effect in this case is not the dynamical motion of the photoemitted electrons through the chiral molecular adlayer, but rather the stabilization of magnetic moments by transverse currents, as demonstrated in the previous section. Hence, once there is a layer of surface moments, the photoemitted electrons are scattered through this layer, and because of the magnetic interaction, different spin species scatter with different rates. This leads to different spin populations reaching the detector, constituting the photoemission CISS effect. With  photon energies of 21 eV, commonly used in photoemission experiments and far in excess of typical molecular ionization energies, it is sufficient to consider this photoemission process in good approximation as a scattering problem.

\begin{figure}
\begin{center}
\includegraphics[width=16.6cm]{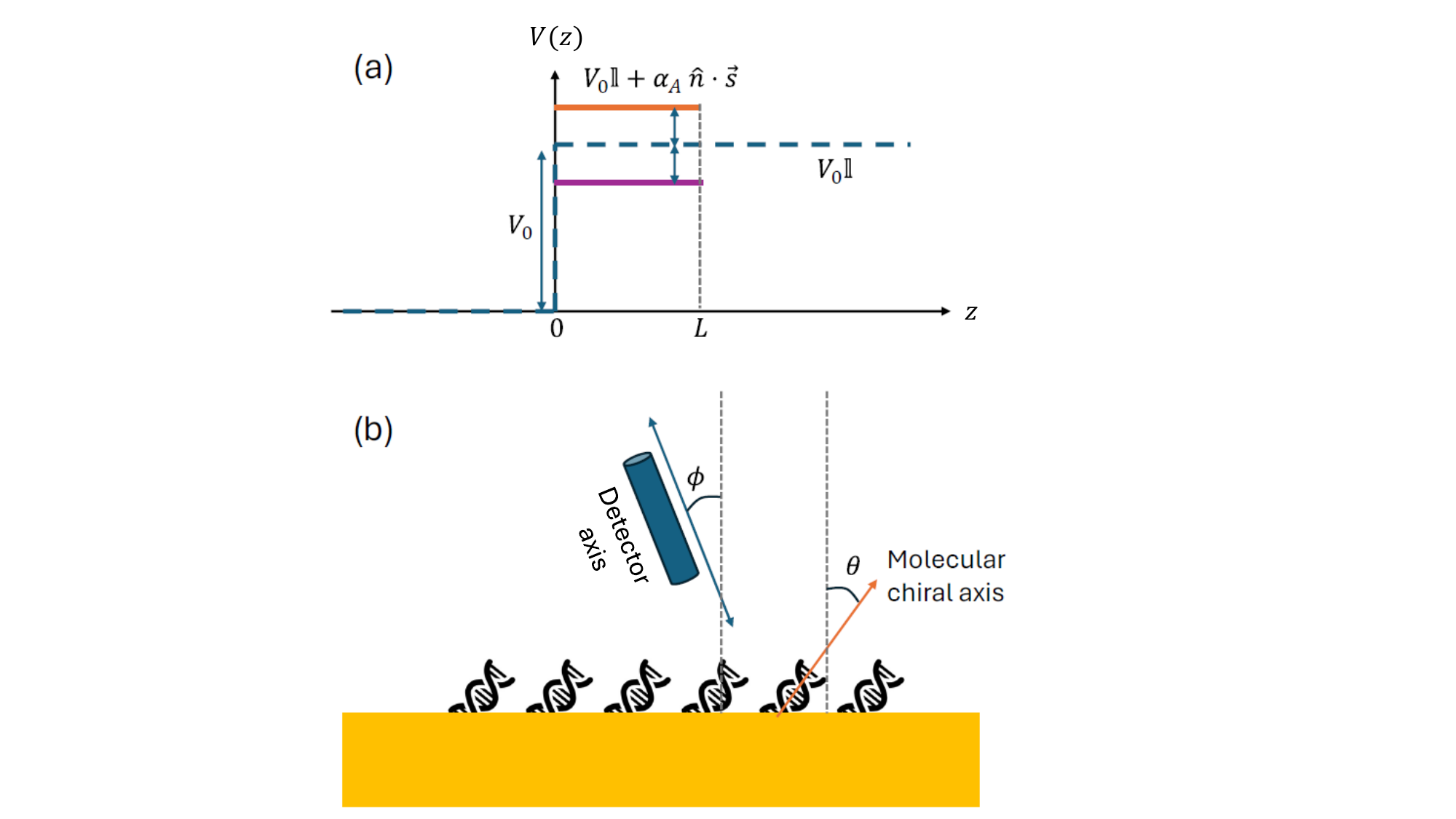}
\caption{(a) The scattering potential for electrons photo-excited from the bulk ($z<0$),  through a scattering region with surface magnetic moments ($0\leq z\leq L$) to the vacuum ($z>L$), and collected by a detector at infinity. (b) Schematic representation of the angles $\theta$ (chiral axis direction) and $\phi$ (the tilt between the detector axis and the surface normal).   }
\label{Fig_schematic2}
\end{center}
\end{figure}

To proceed, we thus construct a scattering theory for photoemitted electrons. The model assumes that at the interface between the metal and molecules, the electrons interact with a layer of molecule-stabilized magnetic moments. This can be described by an interaction Hamiltonian of the form 
\beq
\mathcal{H}_{M}=\alpha_A \hat{n}\cdot \vec{s}~~,
\eeq
where $\alpha_A$ is the electrode SO interaction, representative of the strength of interaction between the surface magnetization and the electron spins (as they pass through the surface), $\hat{n}$ is the molecular chiral axis, and $\vec{s}=\{\sigma_x,\sigma_y,\sigma_z\}$ is the unit-less spin vector. With this interaction in mind, we now construct a 1D scattering problem for electrons scattered from the bulk ($z<0$), through a scattering region affected by the surface magnetic moments ($0\leq z\leq L$) to the vacuum ($z>L$), and collected by a detector at infinity. The potential in $2\times 2$ spinor-space is of the form
\beq
V(z)=
\begin{cases}
0,~z<0~(\mathrm{region~I })\\
V_0 \mathbb{1} +\alpha_A \hat{n}\cdot \vec{s},~0\leq z \leq L ~(\mathrm{region~II})\\
V_0 \mathbb{1}, ~z>L ~(\mathrm{region~III})
\end{cases}
\eeq where $\mathbb{1}$ is the ${2\times2}$ unit matrix. $V_0$ is the vacuum level at the surface, measured from the bottom of the electrode band. The potential is schematically depicted in Fig.~\ref{Fig_schematic2}(a) and for simplicity ignores the far-field evolution of the vacuum level well above the surface \cite{monti2010influence}.

The eigenvalues of the rotation matrix of the spinor term $V_0 \mathbb{1} +\alpha_A \hat{n}\cdot \vec{s}$ can easily be found, and in terms of a standard basis pointing in the direction of the surface normal are simply $|+\rangle=\cos(\frac{\theta}{2})|1\rangle+\sin(\frac{\theta}{2})|-1\rangle$, $|-\rangle=\sin(\frac{\theta}{2})|1\rangle+\cos(\frac{\theta}{2})|-1\rangle$, with eigenvalues $\varepsilon_0=V_0\pm \alpha_A$. Here, $\theta$ is the tilt angle of the chiral axis with respect to the surface, see Fig.~\ref{Fig_schematic2}(b), and $|1\rangle= \tvect{1}{0},|-1\rangle= \tvect{0}{1}$ represent spinors in the z-direction normal to the surface. We note that these eigenvectors only depend on the chiral axis and are independent of the magnetization at the interface.

This implies that the scattering spinor wave-function in region II is 
\beq
\Psi_{II}=\mathcal{R}\left(\frac{\theta}{2}\right) \begin{pmatrix}
           a_1 e^{i k_+ x}+b_1 e^{-i k_+ x} \\
           a_2 e^{i k_- x}+b_2 e^{-i k_- x}
         \end{pmatrix}~~,
\label{PsiII}\eeq
where $\mathcal{R} \left(\frac{\theta}{2}\right)$ is a 2D rotation matrix with angle $\frac{\theta}{2}$, and $k_\pm = \sqrt{\frac{2m (E-V_0\pm \alpha_A)}{\hbar^2}}$ are the momenta, with $E$ the electron energy given by $E=E_f+E_{\mathrm{kin}}$, where $m$ the electron mass, and $E_{\mathrm{kin}}$ is the so-called electron kinetic energy (i.e. electron energy above the vacuum level). The experimental photon energy $\hbar \omega$ is equal to the total electron energy $E$ in this model.  

Considering the fact that in the bulk there is no preferred spin-direction, without loss of generality the scattering wave function in region I can be written as 

\beq
\Psi_{I}=\mathcal{R}\left(\frac{\theta}{2}\right) \left[
\begin{pmatrix}
           1 \\
            1
         \end{pmatrix}e^{i k x} +
\begin{pmatrix}
           r_1 \\
            r_2
         \end{pmatrix}e^{-i k x} \right]
 ~~,
\eeq
where $k = \sqrt{\frac{2m E}{\hbar^2}}$  and $r_{1,2}$ are the reflection amplitudes for the two spin channels. 

Finally, the wave-function in region III is defined by the spin-discrimination axis of the detector. If we assume that the detector is tilted by an angle $\phi$ with respect to the sample surface normal (see Fig.~\ref{Fig_schematic2}(b)), then we have 
\beq
\Psi_{I}=\mathcal{R}(\phi) \begin{pmatrix}
           t_1 \\
            t_2
         \end{pmatrix}e^{i k_1 x}~~,
\eeq
where $k_1 = \sqrt{\frac{2m (E-V_0)}{\hbar^2}}$  and $t_1,t_2$ are the transmission amplitudes for the two spin channels, measured in the axis of the detector. 

The scattering wave-functions are easily found from continuity conditions. The results can be expressed analytically. Defining 
\beq
t_\pm=\frac{2 i k^2 k_\pm e^{-i k_1 L}}{\left(k k_1+k_\pm^2\right) \sin (L k_\pm)+i \left(k+k_1\right) k_\pm \cos (L k_\pm)}~~,
\label{t_pm}
\eeq
and reminding the readers that $k = \sqrt{\frac{2m E}{\hbar^2}},~k_1 = \sqrt{\frac{2m (E-V_0)}{\hbar^2}}, k_{\pm}=\sqrt{\frac{2m (E-V_0\pm\alpha_A)}{\hbar^2}}$,
one obtains a simple form for the transmission amplitudes, $t_1 = t_+ \cos(\phi-\frac{\theta}{2})-t_- \sin(\phi-\frac{\theta}{2})$ and ~$t_2 = t_+ \sin(\phi-\frac{\theta}{2})+t_- \cos(\phi-\frac{\theta}{2})$. The resulting transmissions for the two channels are thus
\begin{eqnarray}
T_1 &=& T_+ \cos^2(\phi-\frac{\theta}{2})+ T_- \sin^2(\phi-\frac{\theta}{2})-2 \mathrm{Re}(t_+t_-)\cos(\phi-\frac{\theta}{2})\sin(\phi-\frac{\theta}{2})\nonumber \\
T_2 &=& T_+ \sin^2(\phi-\frac{\theta}{2})+ T_- \cos^2(\phi-\frac{\theta}{2})+2 \mathrm{Re}(t_+t_-)\cos(\phi-\frac{\theta}{2})\sin(\phi-\frac{\theta}{2})~~, 
\end{eqnarray}
where $T_+=|t_+|^2,~T_-=|t_-|^2$. We again stress that these transmissions represent electrons with spins parallel ($T_1$) or anti-parallel ($T_2$) to the detector axis.

The signal obtained from the detector for the different spins, $I_{\uparrow,\downarrow}$ will be proportional to the transmission of the two spin-channels, $T_{1,2}$ respectively. We define the "CISS signal" for photoemission as the difference in number of electrons scattered to the spin-dependent detector between the two spin-species, proportional to $\Delta T$, which is found to be 
\beq 
\Delta T=T_1-T_2=(T_+-T_-)\cos(2\phi-\theta)-2\mathrm{Re}(t_+ t_-) \sin(2\phi-\theta)~,
\label{T1T2}\eeq
where $T_\pm=|t_\pm|^2$, and $t_\pm$ are defined by Eq.~\ref{t_pm}. 

Eq.~\ref{T1T2} readily allows us to examine the CISS polarization, $P_{CISS}=\frac{\Delta T}{T_1+T_2}$, for different electron kinetic energies and different lengths $L$. Notably, $L$ (defined in Fig.~\ref{Fig_schematic2}) is not the length of the molecule per se, but an effective length which ultimately represents the solenoid field at the interface. We expect it to be proportional to the molecular length if, e.g., the surface magnetization is inherited into the molecule, though these details are subject of future investigations. 

\begin{figure}
\begin{center}
\includegraphics[width=14.6cm]{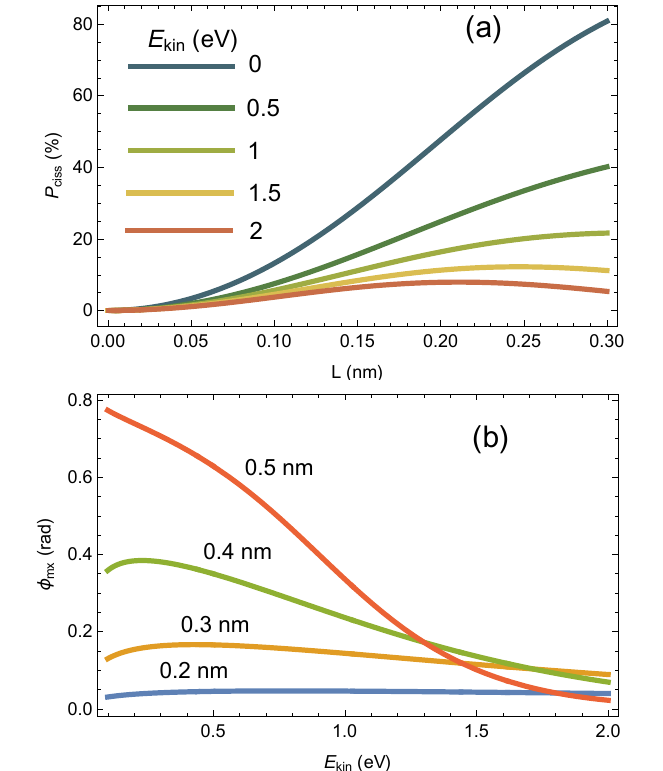}
\caption{(a) CISS polarization, $P_{CISS}$ as a function of scattering region length, $L$, for different electron kinetic energy (i.e. electron energy above vacuum level).  
(b) Angle of maximal CISS effect, $\phi_{mx}$, as a function of kinetic energy $E_{\mathrm{kin}}$, i.e. electron energy above vacuum level, for different scattering region lengths $L=0.2,0.3,0.4,0.5$nm. See text for other numerical parameters. }
\label{phimx}
\end{center}
\end{figure}

In Fig.~\ref{phimx}(a) the CISS polarization, $P_{CISS}$, is plotted as a function of $L$ for different electron kinetic energies, where we take $V_0=5.3$eV as the Au vacuum level with respect to the Fermi level 
and $\alpha_A=0.5$eV \cite{alwan2021spinterface}. As can be seen, the polarization increases with $L$. This effect is most pronounced for slow electrons and is reminiscent of some early studies of the photoemission CISS effect and its dependence on molecular length \cite{Gohler11}.  

Eq.~\ref{T1T2} further provides a directly experimentally measurable prediction for a system in which the spin detector axis can be aligned with respect to the metallic surface. Consider first the case where $\theta=0$, either because the molecular chiral axis is pointing directly normal to the surface, or, as is more likely, if there is some surface disorder and the contributions in the surface $x-y$ plane average to zero. In this case, it is easy to see that the maximal CISS signal $\Delta T$ is {\sl not} at $\phi=0$, but rather at \beq \phi_{mx}=\frac{1}{2}\tan^{-1}\left(\frac{T_+-T_-} {2\mathrm{Re}(t_+ t_-)}\right)~.\eeq
Both the CISS signal $\Delta T$ and the polarization $P_{CISS}=\frac{\Delta T}{T_1+T_2}$ will exhibit a maximum at a tilt angle $\phi_{mx}$, since one can easily verify that $T_1+T_2=T_++T_-$, and thus the denominator of $P$ has no angular dependence.

In Fig.~\ref{phimx}(b) we plot the angle of maximal CISS signal, $\phi_{mx}$, as a function of the electron kinetic energy $E_\mathrm{kin}$ with the scattering region length taken to be $L=0.1,0.2,0.3,0.4$nm (other parameters are same as in Fig.~\ref{phimx}(a)). For short scattering distances, $\phi_{mx}$ is essentially energy-independent, but the dependence becomes more pronounced as the scattering region length increases. 

Finally, if the photoemission CISS experiment is performed with a well-defined chiral axis, e.g. by controlling the tilt angle $\theta$ of the self-assembled monolayer of chiral molecules, then, from Eq.~\ref{T1T2}, $\phi_{mx}$ is predicted to shift by an angle $\theta/2$. By examining these predictions, probing the photoemission CISS effect at different illumination wavelength, detector tilt angle, and different chiral axis, can thus shed new light on the CISS mechanism, and such experiments constitute a direct test of this and other models for CISS.

\section{Summary and Discussion}

To briefly summarize, we have shown that the spinterface mechanism \cite{alwan2021spinterface,alwan2024role,alwan2023temperature,dubi2022spinterface} can account for the dynamical formation of surface magnetic moments and time-reversal symmetry breaking due to displacement currents when chiral molecules adsorb on a metallic surface. The surface magnetic moments are stabilized parallel to the  the direction of the molecular "effective solenoid" chiral axis. We further show with a simple scattering model of photoemission that the photoemission CISS effect may indeed stem from magnetization of the interface between a metallic substrate and chiral molecules. The model exhibits purely spinterface-induced length-dependence, long considered a hallmark fingerprint of CISS in photoemission. Remarkably, our model predicts strong deviations of maximal spin polarization away from surface normal or chiral axis, particularly pronounced for slow electrons. At least for photoemission, the model developed here suggests that CISS need not be generated by chiral scattering in an adsorbed molecule but is rather caused by the spinterface. Importantly, photoemission CISS does not require helical molecules but is rather more general.

Our model has important ramifications that may be tested experimentally. The fact that spin polarization should be sought along the chiral axis rather than the surface normal is perhaps to be expected, but has been largely ignored in the CISS literature. It may indeed be relevant for the interpretation of  previous experimental results, as well as for the design of future experiments. Most obviously, we envisage the possibility to directly test our predictions with spin- and angle-resolved photoemission spectroscopy (SARPES) \cite{baljozovic2023spin}, where the signal is highly sensitive to the magnetic propeties of the surface, and thus detailed control over the tilt angle and hence the chiral axis of the molecules on the surface will dramatically affect the SARPES signal. 

We expect that similar considerations may also apply to transport CISS experiments. Here, the ferromagnetic (FM) electrode is typically magnetized parallel or anti-parallel to the direction of current, which is perpendicular to the electrode surface. In an experiment in which the FM magnetization direction can be rotated and in which there is a well-defined molecular chiral axis, maximal spin polarization will be measured when the FM magnetization is pointing parallel to the chiral axis, and not along the direction of current. In order to test this prediction, a highly ordered molecular layer is of utmost importance, since the in-plane component of the surface moments will be averaged to zero otherwise.

An intriguing extension of CISS concerns the measurement of in-plane current which does not necessarily flow through the chiral molecules on the surface (see e.g. \cite{kondou2022chirality}). A possible explanation for the results of Ref. ~\cite{kondou2022chirality} is the scattering of electrons off surface moments which are stabilized by the presence of chiral molecules on the surface. Controlling the surface angle of the chiral molecules would then change the direction of the surface moments, and hence the scattering direction. 

The prediction that observed spin polarization is maximized at angles away from the chiral axis given sufficiently long molecules and low enough kinetic energies is intriguing. Fundamentally, it results from a combination of the direction of the stabilized magnetization at the interface and a projection of the spin axis onto the spin detector axis. Since the photoemission is at its core a scattering problem, amplitudes for the different spin channels interfere in the resulting transmission, giving rise to the non-trivial angle dependence of  $\phi_{mx}$. Note also that $T_{1,2}$ can be larger than unity, because the scattering region acts to convert between the spin-species, and the mathematical limitation here is $T_1+T_2 \leq 2$. 

We finish by emphasizing that the notion of a fluctuating surface magnetization is supported by numerous observations e.g. in gold nanoparticles or bare Au surfaces, obtained under different conditions and without adsorbed chiral molecules, and is directly related to the strong spin-orbit (SO) interactions in gold. \cite{trudel2011unexpected, li2011intrinsic, sato2015magnetic, nealon2012magnetism, agrachev2017magnetic, henk2003spin}. As a result, one may expect the spinterface mechanism to be rather general and to manifest not just in transmission and photoemission CISS experiments, but any experiment that is fundamentally interface sensitive.

\section{Acknowledgments}
The authors are grateful to Prof. A. Sharoni for many fruitful and illuminating discussions. OLAM gratefully acknowledges support by the Air Force Office of Scientific Research under award FA9550-21-1-0219.

\bibliography{refs}

\providecommand{\latin}[1]{#1}
\makeatletter
\providecommand{\doi}
  {\begingroup\let\do\@makeother\dospecials
  \catcode`\{=1 \catcode`\}=2 \doi@aux}
\providecommand{\doi@aux}[1]{\endgroup\texttt{#1}}
\makeatother
\providecommand*\mcitethebibliography{\thebibliography}
\csname @ifundefined\endcsname{endmcitethebibliography}  {\let\endmcitethebibliography\endthebibliography}{}
\begin{mcitethebibliography}{69}
\providecommand*\natexlab[1]{#1}
\providecommand*\mciteSetBstSublistMode[1]{}
\providecommand*\mciteSetBstMaxWidthForm[2]{}
\providecommand*\mciteBstWouldAddEndPuncttrue
  {\def\EndOfBibitem{\unskip.}}
\providecommand*\mciteBstWouldAddEndPunctfalse
  {\let\EndOfBibitem\relax}
\providecommand*\mciteSetBstMidEndSepPunct[3]{}
\providecommand*\mciteSetBstSublistLabelBeginEnd[3]{}
\providecommand*\EndOfBibitem{}
\mciteSetBstSublistMode{f}
\mciteSetBstMaxWidthForm{subitem}{(\alph{mcitesubitemcount})}
\mciteSetBstSublistLabelBeginEnd
  {\mcitemaxwidthsubitemform\space}
  {\relax}
  {\relax}

\bibitem[Aiello \latin{et~al.}(2022)Aiello, Abendroth, Abbas, Afanasev, Agarwal, Banerjee, Beratan, Belling, Berche, Botana, \latin{et~al.} others]{aiello2022chirality}
Aiello,~C.~D.; Abendroth,~J.~M.; Abbas,~M.; Afanasev,~A.; Agarwal,~S.; Banerjee,~A.~S.; Beratan,~D.~N.; Belling,~J.~N.; Berche,~B.; Botana,~A.; others A Chirality-Based Quantum Leap. \emph{ACS nano} \textbf{2022}, \emph{16}, 4989--5035\relax
\mciteBstWouldAddEndPuncttrue
\mciteSetBstMidEndSepPunct{\mcitedefaultmidpunct}
{\mcitedefaultendpunct}{\mcitedefaultseppunct}\relax
\EndOfBibitem
\bibitem[Xu and Mi(2023)Xu, and Mi]{xu2023chiral}
Xu,~Y.; Mi,~W. Chiral-induced spin selectivity in biomolecules, hybrid organic inorganic perovskites and inorganic materials: a comprehensive review on recent progress. \emph{Materials Horizons} \textbf{2023}, \emph{10}, 1924--1955\relax
\mciteBstWouldAddEndPuncttrue
\mciteSetBstMidEndSepPunct{\mcitedefaultmidpunct}
{\mcitedefaultendpunct}{\mcitedefaultseppunct}\relax
\EndOfBibitem
\bibitem[Aragon{\`e}s \latin{et~al.}(2022)Aragon{\`e}s, Aravena, Ugalde, Medina, Gutierrez, Ruiz, Mujica, and D{\'\i}ez-P{\'e}rez]{aragones2022magnetoresistive}
Aragon{\`e}s,~A.~C.; Aravena,~D.; Ugalde,~J.~M.; Medina,~E.; Gutierrez,~R.; Ruiz,~E.; Mujica,~V.; D{\'\i}ez-P{\'e}rez,~I. Magnetoresistive Single-Molecule Junctions: the Role of the Spinterface and the CISS Effect. \emph{Israel Journal of Chemistry} \textbf{2022}, e202200090\relax
\mciteBstWouldAddEndPuncttrue
\mciteSetBstMidEndSepPunct{\mcitedefaultmidpunct}
{\mcitedefaultendpunct}{\mcitedefaultseppunct}\relax
\EndOfBibitem
\bibitem[Bloom \latin{et~al.}(2024)Bloom, Paltiel, Naaman, and Waldeck]{Bloom2024chiral}
Bloom,~B.~P.; Paltiel,~Y.; Naaman,~R.; Waldeck,~D.~H. Chiral Induced Spin Selectivity. \emph{Chemical Reviews} \textbf{2024}, \emph{124}, 1950--1991, PMID: 38364021\relax
\mciteBstWouldAddEndPuncttrue
\mciteSetBstMidEndSepPunct{\mcitedefaultmidpunct}
{\mcitedefaultendpunct}{\mcitedefaultseppunct}\relax
\EndOfBibitem
\bibitem[Naaman and Waldeck(2012)Naaman, and Waldeck]{Naaman12}
Naaman,~R.; Waldeck,~D.~H. Chiral-Induced Spin Selectivity Effect. \emph{The Journal of Physical Chemistry Letters} \textbf{2012}, \emph{3}, 2178--2187, PMID: 26295768\relax
\mciteBstWouldAddEndPuncttrue
\mciteSetBstMidEndSepPunct{\mcitedefaultmidpunct}
{\mcitedefaultendpunct}{\mcitedefaultseppunct}\relax
\EndOfBibitem
\bibitem[Naaman and Waldeck(2015)Naaman, and Waldeck]{Naaman15}
Naaman,~R.; Waldeck,~D.~H. Spintronics and Chirality: Spin Selectivity in Electron Transport Through Chiral Molecules. \emph{Annual Review of Physical Chemistry} \textbf{2015}, \emph{66}, 263--281, PMID: 25622190\relax
\mciteBstWouldAddEndPuncttrue
\mciteSetBstMidEndSepPunct{\mcitedefaultmidpunct}
{\mcitedefaultendpunct}{\mcitedefaultseppunct}\relax
\EndOfBibitem
\bibitem[Naaman \latin{et~al.}(2019)Naaman, Paltiel, and Waldeck]{naaman2019chiral}
Naaman,~R.; Paltiel,~Y.; Waldeck,~D.~H. Chiral molecules and the electron spin. \emph{Nature Reviews Chemistry} \textbf{2019}, \emph{3}, 250--260\relax
\mciteBstWouldAddEndPuncttrue
\mciteSetBstMidEndSepPunct{\mcitedefaultmidpunct}
{\mcitedefaultendpunct}{\mcitedefaultseppunct}\relax
\EndOfBibitem
\bibitem[Naaman \latin{et~al.}(2020)Naaman, Paltiel, and Waldeck]{naaman2020chiral}
Naaman,~R.; Paltiel,~Y.; Waldeck,~D. Chiral molecules and the spin selectivity effect. \emph{The journal of physical chemistry letters} \textbf{2020}, \emph{11}, 3660--3666\relax
\mciteBstWouldAddEndPuncttrue
\mciteSetBstMidEndSepPunct{\mcitedefaultmidpunct}
{\mcitedefaultendpunct}{\mcitedefaultseppunct}\relax
\EndOfBibitem
\bibitem[M{\"o}llers \latin{et~al.}(2022)M{\"o}llers, G{\"o}hler, and Zacharias]{mollers2022chirality}
M{\"o}llers,~P.~V.; G{\"o}hler,~B.; Zacharias,~H. Chirality Induced Spin Selectivity--the Photoelectron View. \emph{Israel Journal of Chemistry} \textbf{2022}, e202200062\relax
\mciteBstWouldAddEndPuncttrue
\mciteSetBstMidEndSepPunct{\mcitedefaultmidpunct}
{\mcitedefaultendpunct}{\mcitedefaultseppunct}\relax
\EndOfBibitem
\bibitem[Kettner \latin{et~al.}(2018)Kettner, Maslyuk, Nurenberg, Seibel, Gutierrez, Cuniberti, Ernst, and Zacharias]{kettner2018chirality}
Kettner,~M.; Maslyuk,~V.~V.; Nurenberg,~D.; Seibel,~J.; Gutierrez,~R.; Cuniberti,~G.; Ernst,~K.-H.; Zacharias,~H. Chirality-dependent electron spin filtering by molecular monolayers of helicenes. \emph{The journal of physical chemistry letters} \textbf{2018}, \emph{9}, 2025--2030\relax
\mciteBstWouldAddEndPuncttrue
\mciteSetBstMidEndSepPunct{\mcitedefaultmidpunct}
{\mcitedefaultendpunct}{\mcitedefaultseppunct}\relax
\EndOfBibitem
\bibitem[Möllers \latin{et~al.}(2022)Möllers, Wei, Salamon, Bartsch, Wende, Waldeck, and Zacharias]{mollers2022spin}
Möllers,~P.~V.; Wei,~J.; Salamon,~S.; Bartsch,~M.; Wende,~H.; Waldeck,~D.~H.; Zacharias,~H. Spin-polarized photoemission from chiral CuO catalyst thin films. \emph{ACS nano} \textbf{2022}, \emph{16}, 12145--12155\relax
\mciteBstWouldAddEndPuncttrue
\mciteSetBstMidEndSepPunct{\mcitedefaultmidpunct}
{\mcitedefaultendpunct}{\mcitedefaultseppunct}\relax
\EndOfBibitem
\bibitem[Gohler \latin{et~al.}(2011)Gohler, Hamelbeck, Markus, Kettner, Hanne, Vager, Naaman, and Zacharias]{Gohler11}
Gohler,~B.; Hamelbeck,~V.; Markus,~T.~Z.; Kettner,~M.; Hanne,~G.~F.; Vager,~Z.; Naaman,~R.; Zacharias,~H. Spin Selectivity in Electron Transmission Through Self-Assembled Monolayers of Double-Stranded DNA. \emph{Science} \textbf{2011}, \emph{331}, 894--897\relax
\mciteBstWouldAddEndPuncttrue
\mciteSetBstMidEndSepPunct{\mcitedefaultmidpunct}
{\mcitedefaultendpunct}{\mcitedefaultseppunct}\relax
\EndOfBibitem
\bibitem[Abendroth \latin{et~al.}(2019)Abendroth, Cheung, Stemer, El~Hadri, Zhao, Fullerton, and Weiss]{abendroth2019spin}
Abendroth,~J.~M.; Cheung,~K.~M.; Stemer,~D.~M.; El~Hadri,~M.~S.; Zhao,~C.; Fullerton,~E.~E.; Weiss,~P.~S. Spin-dependent ionization of chiral molecular films. \emph{Journal of the American Chemical Society} \textbf{2019}, \emph{141}, 3863--3874\relax
\mciteBstWouldAddEndPuncttrue
\mciteSetBstMidEndSepPunct{\mcitedefaultmidpunct}
{\mcitedefaultendpunct}{\mcitedefaultseppunct}\relax
\EndOfBibitem
\bibitem[Badala~Viswanatha \latin{et~al.}(2022)Badala~Viswanatha, Stöckl, Arnoldi, Becker, Aeschlimann, and Stadtmüller]{badala2022vectorial}
Badala~Viswanatha,~C.; Stöckl,~J.; Arnoldi,~B.; Becker,~S.; Aeschlimann,~M.; Stadtmüller,~B. Vectorial electron spin filtering by an all-chiral metal--molecule heterostructure. \emph{The Journal of Physical Chemistry Letters} \textbf{2022}, \emph{13}, 6244--6249\relax
\mciteBstWouldAddEndPuncttrue
\mciteSetBstMidEndSepPunct{\mcitedefaultmidpunct}
{\mcitedefaultendpunct}{\mcitedefaultseppunct}\relax
\EndOfBibitem
\bibitem[Ni{\~n}o \latin{et~al.}(2014)Ni{\~n}o, Kowalik, Luque, Arvanitis, Miranda, and de~Miguel]{nino2014enantiospecific}
Ni{\~n}o,~M.~{\'A}.; Kowalik,~I.~A.; Luque,~F.~J.; Arvanitis,~D.; Miranda,~R.; de~Miguel,~J.~J. Enantiospecific spin polarization of electrons photoemitted through layers of homochiral organic molecules. \emph{Advanced Materials} \textbf{2014}, \emph{26}, 7474--7479\relax
\mciteBstWouldAddEndPuncttrue
\mciteSetBstMidEndSepPunct{\mcitedefaultmidpunct}
{\mcitedefaultendpunct}{\mcitedefaultseppunct}\relax
\EndOfBibitem
\bibitem[Yang \latin{et~al.}(2023)Yang, Li, Zhou, Guo, Jia, Liu, Houk, Dubi, and Guo]{yang2023}
Yang,~C.; Li,~Y.; Zhou,~S.; Guo,~Y.; Jia,~C.; Liu,~Z.; Houk,~K.~N.; Dubi,~Y.; Guo,~X. Real-time monitoring of reaction stereochemistry through single-molecule observations of chirality-induced spin selectivity. \emph{Nature Chemistry} \textbf{2023}, \emph{15}, 972--979\relax
\mciteBstWouldAddEndPuncttrue
\mciteSetBstMidEndSepPunct{\mcitedefaultmidpunct}
{\mcitedefaultendpunct}{\mcitedefaultseppunct}\relax
\EndOfBibitem
\bibitem[Ben~Dor \latin{et~al.}(2017)Ben~Dor, Yochelis, Radko, Vankayala, Capua, Capua, Yang, Baczewski, Parkin, Naaman, \latin{et~al.} others]{ben2017magnetization}
Ben~Dor,~O.; Yochelis,~S.; Radko,~A.; Vankayala,~K.; Capua,~E.; Capua,~A.; Yang,~S.-H.; Baczewski,~L.~T.; Parkin,~S. S.~P.; Naaman,~R.; others Magnetization switching in ferromagnets by adsorbed chiral molecules without current or external magnetic field. \emph{Nature communications} \textbf{2017}, \emph{8}, 14567\relax
\mciteBstWouldAddEndPuncttrue
\mciteSetBstMidEndSepPunct{\mcitedefaultmidpunct}
{\mcitedefaultendpunct}{\mcitedefaultseppunct}\relax
\EndOfBibitem
\bibitem[Koplovitz \latin{et~al.}(2019)Koplovitz, Leitus, Ghosh, Bloom, Yochelis, Rotem, Vischio, Striccoli, Fanizza, Naaman, \latin{et~al.} others]{koplovitz2019single}
Koplovitz,~G.; Leitus,~G.; Ghosh,~S.; Bloom,~B.~P.; Yochelis,~S.; Rotem,~D.; Vischio,~F.; Striccoli,~M.; Fanizza,~E.; Naaman,~R.; others Single domain 10 nm ferromagnetism imprinted on superparamagnetic nanoparticles using chiral molecules. \emph{Small} \textbf{2019}, \emph{15}, 1804557\relax
\mciteBstWouldAddEndPuncttrue
\mciteSetBstMidEndSepPunct{\mcitedefaultmidpunct}
{\mcitedefaultendpunct}{\mcitedefaultseppunct}\relax
\EndOfBibitem
\bibitem[Metzger \latin{et~al.}(2020)Metzger, Mishra, Bloom, Goren, Neubauer, Shmul, Wei, Yochelis, Tassinari, Fontanesi, \latin{et~al.} others]{metzger2020electron}
Metzger,~T.~S.; Mishra,~S.; Bloom,~B.~P.; Goren,~N.; Neubauer,~A.; Shmul,~G.; Wei,~J.; Yochelis,~S.; Tassinari,~F.; Fontanesi,~C.; others The electron spin as a chiral reagent. \emph{Angewandte Chemie} \textbf{2020}, \emph{132}, 1670--1675\relax
\mciteBstWouldAddEndPuncttrue
\mciteSetBstMidEndSepPunct{\mcitedefaultmidpunct}
{\mcitedefaultendpunct}{\mcitedefaultseppunct}\relax
\EndOfBibitem
\bibitem[Tassinari \latin{et~al.}(2018)Tassinari, Jayarathna, Kantor-Uriel, Davis, Varade, Achim, and Naaman]{tassinari2018chirality}
Tassinari,~F.; Jayarathna,~D.~R.; Kantor-Uriel,~N.; Davis,~K.~L.; Varade,~V.; Achim,~C.; Naaman,~R. Chirality dependent charge transfer rate in oligopeptides. \emph{Advanced Materials} \textbf{2018}, \emph{30}, 1706423\relax
\mciteBstWouldAddEndPuncttrue
\mciteSetBstMidEndSepPunct{\mcitedefaultmidpunct}
{\mcitedefaultendpunct}{\mcitedefaultseppunct}\relax
\EndOfBibitem
\bibitem[Abendroth \latin{et~al.}(2017)Abendroth, Nakatsuka, Ye, Kim, Fullerton, Andrews, and Weiss]{abendroth2017analyzing}
Abendroth,~J.~M.; Nakatsuka,~N.; Ye,~M.; Kim,~D.; Fullerton,~E.~E.; Andrews,~A.~M.; Weiss,~P.~S. Analyzing spin selectivity in DNA-mediated charge transfer via fluorescence microscopy. \emph{ACS nano} \textbf{2017}, \emph{11}, 7516--7526\relax
\mciteBstWouldAddEndPuncttrue
\mciteSetBstMidEndSepPunct{\mcitedefaultmidpunct}
{\mcitedefaultendpunct}{\mcitedefaultseppunct}\relax
\EndOfBibitem
\bibitem[Mishra \latin{et~al.}(2024)Mishra, Bowes, Majumder, Hollingsworth, Htoon, and Jones]{mishra2024inducing}
Mishra,~S.; Bowes,~E.~G.; Majumder,~S.; Hollingsworth,~J.~A.; Htoon,~H.; Jones,~A.~C. Inducing Circularly Polarized Single-Photon Emission via Chiral-Induced Spin Selectivity. \emph{ACS nano} \textbf{2024}, \emph{18}, 8663--8672\relax
\mciteBstWouldAddEndPuncttrue
\mciteSetBstMidEndSepPunct{\mcitedefaultmidpunct}
{\mcitedefaultendpunct}{\mcitedefaultseppunct}\relax
\EndOfBibitem
\bibitem[Theiler \latin{et~al.}(2023)Theiler, Ritz, Hofmann, and Stemmer]{theiler2023detection}
Theiler,~P.~M.; Ritz,~C.; Hofmann,~R.; Stemmer,~A. Detection of a Chirality-Induced Spin Selective Quantum Capacitance in $\alpha$-Helical Peptides. \emph{Nano Letters} \textbf{2023}, \emph{23}, 8280--8287\relax
\mciteBstWouldAddEndPuncttrue
\mciteSetBstMidEndSepPunct{\mcitedefaultmidpunct}
{\mcitedefaultendpunct}{\mcitedefaultseppunct}\relax
\EndOfBibitem
\bibitem[Nguyen \latin{et~al.}(2024)Nguyen, Salvan, Hellwig, Paltiel, Baczewski, and Tegenkamp]{Nguyen2024}
Nguyen,~T. N.~H.; Salvan,~G.; Hellwig,~O.; Paltiel,~Y.; Baczewski,~L.~T.; Tegenkamp,~C. The mechanism of the molecular CISS effect in chiral nano-junctions. \emph{Chem. Sci.} \textbf{2024}, --\relax
\mciteBstWouldAddEndPuncttrue
\mciteSetBstMidEndSepPunct{\mcitedefaultmidpunct}
{\mcitedefaultendpunct}{\mcitedefaultseppunct}\relax
\EndOfBibitem
\bibitem[Evers \latin{et~al.}(2020)Evers, Koryt\'ar, Tewari, and van Ruitenbeek]{Evers20}
Evers,~F.; Koryt\'ar,~R.; Tewari,~S.; van Ruitenbeek,~J.~M. Advances and challenges in single-molecule electron transport. \emph{Rev. Mod. Phys.} \textbf{2020}, \emph{92}, 035001\relax
\mciteBstWouldAddEndPuncttrue
\mciteSetBstMidEndSepPunct{\mcitedefaultmidpunct}
{\mcitedefaultendpunct}{\mcitedefaultseppunct}\relax
\EndOfBibitem
\bibitem[Liu and Weiss(2023)Liu, and Weiss]{liu2023spin}
Liu,~T.; Weiss,~P.~S. Spin Polarization in Transport Studies of Chirality-Induced Spin Selectivity. \emph{ACS nano} \textbf{2023}, \emph{17}, 19502--19507\relax
\mciteBstWouldAddEndPuncttrue
\mciteSetBstMidEndSepPunct{\mcitedefaultmidpunct}
{\mcitedefaultendpunct}{\mcitedefaultseppunct}\relax
\EndOfBibitem
\bibitem[Tirion and van Wees(2024)Tirion, and van Wees]{tirion2024mechanism}
Tirion,~S.~H.; van Wees,~B.~J. Mechanism for Electrostatically Generated Magnetoresistance in Chiral Systems without Spin-Dependent Transport. \emph{ACS nano} \textbf{2024}, \emph{18}, 6028--6037\relax
\mciteBstWouldAddEndPuncttrue
\mciteSetBstMidEndSepPunct{\mcitedefaultmidpunct}
{\mcitedefaultendpunct}{\mcitedefaultseppunct}\relax
\EndOfBibitem
\bibitem[Medina \latin{et~al.}(2012)Medina, L{\'o}pez, Ratner, and Mujica]{medina2012chiral}
Medina,~E.; L{\'o}pez,~F.; Ratner,~M.~A.; Mujica,~V. Chiral molecular films as electron polarizers and polarization modulators. \emph{Europhysics Letters} \textbf{2012}, \emph{99}, 17006\relax
\mciteBstWouldAddEndPuncttrue
\mciteSetBstMidEndSepPunct{\mcitedefaultmidpunct}
{\mcitedefaultendpunct}{\mcitedefaultseppunct}\relax
\EndOfBibitem
\bibitem[Eremko and Loktev(2013)Eremko, and Loktev]{eremko2013spin}
Eremko,~A.; Loktev,~V. Spin sensitive electron transmission through helical potentials. \emph{Physical Review B—Condensed Matter and Materials Physics} \textbf{2013}, \emph{88}, 165409\relax
\mciteBstWouldAddEndPuncttrue
\mciteSetBstMidEndSepPunct{\mcitedefaultmidpunct}
{\mcitedefaultendpunct}{\mcitedefaultseppunct}\relax
\EndOfBibitem
\bibitem[Ghazaryan \latin{et~al.}(2020)Ghazaryan, Paltiel, and Lemeshko]{Ghazaryan20}
Ghazaryan,~A.; Paltiel,~Y.; Lemeshko,~M. Analytic Model of Chiral-Induced Spin Selectivity. \emph{The Journal of Physical Chemistry C} \textbf{2020}, \emph{124}, 11716--11721\relax
\mciteBstWouldAddEndPuncttrue
\mciteSetBstMidEndSepPunct{\mcitedefaultmidpunct}
{\mcitedefaultendpunct}{\mcitedefaultseppunct}\relax
\EndOfBibitem
\bibitem[Varela \latin{et~al.}(2013)Varela, Medina, Lopez, and Mujica]{varela2013inelastic}
Varela,~S.; Medina,~E.; Lopez,~F.; Mujica,~V. Inelastic electron scattering from a helical potential: transverse polarization and the structure factor in the single scattering approximation. \emph{Journal of Physics: Condensed Matter} \textbf{2013}, \emph{26}, 015008\relax
\mciteBstWouldAddEndPuncttrue
\mciteSetBstMidEndSepPunct{\mcitedefaultmidpunct}
{\mcitedefaultendpunct}{\mcitedefaultseppunct}\relax
\EndOfBibitem
\bibitem[Fransson(2021)]{fransson2021charge}
Fransson,~J. Charge Redistribution and Spin Polarization Driven by Correlation Induced Electron Exchange in Chiral Molecules. \emph{Nano Letters} \textbf{2021}, \emph{21}, 3026--3032, PMID: 33759530\relax
\mciteBstWouldAddEndPuncttrue
\mciteSetBstMidEndSepPunct{\mcitedefaultmidpunct}
{\mcitedefaultendpunct}{\mcitedefaultseppunct}\relax
\EndOfBibitem
\bibitem[Shiranzaei \latin{et~al.}(2023)Shiranzaei, Kalhöfer, and Fransson]{shiranzaei2023emergent}
Shiranzaei,~M.; Kalhöfer,~S.; Fransson,~J. Emergent magnetism as a cooperative effect of interactions and reservoir. \emph{The Journal of Physical Chemistry Letters} \textbf{2023}, \emph{14}, 5119--5126\relax
\mciteBstWouldAddEndPuncttrue
\mciteSetBstMidEndSepPunct{\mcitedefaultmidpunct}
{\mcitedefaultendpunct}{\mcitedefaultseppunct}\relax
\EndOfBibitem
\bibitem[Fransson(2023)]{fransson2023vibrationally}
Fransson,~J. Vibrationally induced magnetism in supramolecular aggregates. \emph{The Journal of Physical Chemistry Letters} \textbf{2023}, \emph{14}, 2558--2564\relax
\mciteBstWouldAddEndPuncttrue
\mciteSetBstMidEndSepPunct{\mcitedefaultmidpunct}
{\mcitedefaultendpunct}{\mcitedefaultseppunct}\relax
\EndOfBibitem
\bibitem[Fransson(2022)]{fransson2022charge}
Fransson,~J. Charge and spin dynamics and enantioselectivity in chiral molecules. \emph{The Journal of Physical Chemistry Letters} \textbf{2022}, \emph{13}, 808--814\relax
\mciteBstWouldAddEndPuncttrue
\mciteSetBstMidEndSepPunct{\mcitedefaultmidpunct}
{\mcitedefaultendpunct}{\mcitedefaultseppunct}\relax
\EndOfBibitem
\bibitem[Alwan and Dubi(2021)Alwan, and Dubi]{alwan2021spinterface}
Alwan,~S.; Dubi,~Y. Spinterface Origin for the Chirality-Induced Spin-Selectivity Effect. \emph{Journal of the American Chemical Society} \textbf{2021}, \emph{143}, 14235--14241\relax
\mciteBstWouldAddEndPuncttrue
\mciteSetBstMidEndSepPunct{\mcitedefaultmidpunct}
{\mcitedefaultendpunct}{\mcitedefaultseppunct}\relax
\EndOfBibitem
\bibitem[Dubi(2022)]{dubi2022spinterface}
Dubi,~Y. Spinterface chirality-induced spin selectivity effect in bio-molecules. \emph{Chemical Science} \textbf{2022}, \emph{13}, 10878--10883\relax
\mciteBstWouldAddEndPuncttrue
\mciteSetBstMidEndSepPunct{\mcitedefaultmidpunct}
{\mcitedefaultendpunct}{\mcitedefaultseppunct}\relax
\EndOfBibitem
\bibitem[Alwan \latin{et~al.}(2023)Alwan, Sarkar, Sharoni, and Dubi]{alwan2023temperature}
Alwan,~S.; Sarkar,~S.; Sharoni,~A.; Dubi,~Y. Temperature-dependence of the chirality-induced spin selectivity effect—Experiments and theory. \emph{Journal of Chemical Physics} \textbf{2023}, \emph{159}, 014106\relax
\mciteBstWouldAddEndPuncttrue
\mciteSetBstMidEndSepPunct{\mcitedefaultmidpunct}
{\mcitedefaultendpunct}{\mcitedefaultseppunct}\relax
\EndOfBibitem
\bibitem[Alwan \latin{et~al.}(2024)Alwan, Sharoni, and Dubi]{alwan2024role}
Alwan,~S.; Sharoni,~A.; Dubi,~Y. Role of Electrode Polarization in the Electron Transport Chirality-Induced Spin-Selectivity Effect. \emph{The Journal of Physical Chemistry C} \textbf{2024}, \emph{128}, 6438--6445\relax
\mciteBstWouldAddEndPuncttrue
\mciteSetBstMidEndSepPunct{\mcitedefaultmidpunct}
{\mcitedefaultendpunct}{\mcitedefaultseppunct}\relax
\EndOfBibitem
\bibitem[Monti(2012)]{monti2012understanding}
Monti,~O.~L. Understanding interfacial electronic structure and charge transfer: an electrostatic perspective. \emph{The Journal of Physical Chemistry Letters} \textbf{2012}, \emph{3}, 2342--2351\relax
\mciteBstWouldAddEndPuncttrue
\mciteSetBstMidEndSepPunct{\mcitedefaultmidpunct}
{\mcitedefaultendpunct}{\mcitedefaultseppunct}\relax
\EndOfBibitem
\bibitem[Lindblad(1976)]{lindblad1976generators}
Lindblad,~G. On the generators of quantum dynamical semigroups. \emph{Communications in Mathematical Physics} \textbf{1976}, \emph{48}, 119--130\relax
\mciteBstWouldAddEndPuncttrue
\mciteSetBstMidEndSepPunct{\mcitedefaultmidpunct}
{\mcitedefaultendpunct}{\mcitedefaultseppunct}\relax
\EndOfBibitem
\bibitem[Pearle(2012)]{pearle2012simple}
Pearle,~P. Simple derivation of the Lindblad equation. \emph{European journal of physics} \textbf{2012}, \emph{33}, 805\relax
\mciteBstWouldAddEndPuncttrue
\mciteSetBstMidEndSepPunct{\mcitedefaultmidpunct}
{\mcitedefaultendpunct}{\mcitedefaultseppunct}\relax
\EndOfBibitem
\bibitem[Breuer and Petruccione(2002)Breuer, and Petruccione]{breuer2002theory}
Breuer,~H.-P.; Petruccione,~F. \emph{The theory of open quantum systems}; OUP Oxford, 2002\relax
\mciteBstWouldAddEndPuncttrue
\mciteSetBstMidEndSepPunct{\mcitedefaultmidpunct}
{\mcitedefaultendpunct}{\mcitedefaultseppunct}\relax
\EndOfBibitem
\bibitem[Manzano(2020)]{manzano2020short}
Manzano,~D. A short introduction to the Lindblad master equation. \emph{Aip Advances} \textbf{2020}, \emph{10}\relax
\mciteBstWouldAddEndPuncttrue
\mciteSetBstMidEndSepPunct{\mcitedefaultmidpunct}
{\mcitedefaultendpunct}{\mcitedefaultseppunct}\relax
\EndOfBibitem
\bibitem[Landau and Lifshitz(1935)Landau, and Lifshitz]{landau1992theory}
Landau,~L.; Lifshitz,~E. On the theory of the dispersion of magnetic permeability in ferromagnetic bodies. \emph{Phys. Z. Sowjetunion} \textbf{1935}, \emph{8}, 101--114\relax
\mciteBstWouldAddEndPuncttrue
\mciteSetBstMidEndSepPunct{\mcitedefaultmidpunct}
{\mcitedefaultendpunct}{\mcitedefaultseppunct}\relax
\EndOfBibitem
\bibitem[Gilbert(2004)]{gilbert2004phenomenological}
Gilbert,~T.~L. A phenomenological theory of damping in ferromagnetic materials. \emph{IEEE transactions on magnetics} \textbf{2004}, \emph{40}, 3443--3449\relax
\mciteBstWouldAddEndPuncttrue
\mciteSetBstMidEndSepPunct{\mcitedefaultmidpunct}
{\mcitedefaultendpunct}{\mcitedefaultseppunct}\relax
\EndOfBibitem
\bibitem[Lakshmanan(2011)]{lakshmanan2011fascinating}
Lakshmanan,~M. The fascinating world of the Landau--Lifshitz--Gilbert equation: an overview. \emph{Philosophical Transactions of the Royal Society A: Mathematical, Physical and Engineering Sciences} \textbf{2011}, \emph{369}, 1280--1300\relax
\mciteBstWouldAddEndPuncttrue
\mciteSetBstMidEndSepPunct{\mcitedefaultmidpunct}
{\mcitedefaultendpunct}{\mcitedefaultseppunct}\relax
\EndOfBibitem
\bibitem[Stiles and Miltat(2006)Stiles, and Miltat]{stiles2006spin}
Stiles,~M.~D.; Miltat,~J. Spin-transfer torque and dynamics. \emph{Spin dynamics in confined magnetic structures III} \textbf{2006}, 225--308\relax
\mciteBstWouldAddEndPuncttrue
\mciteSetBstMidEndSepPunct{\mcitedefaultmidpunct}
{\mcitedefaultendpunct}{\mcitedefaultseppunct}\relax
\EndOfBibitem
\bibitem[Dubi and Di~Ventra(2009)Dubi, and Di~Ventra]{dubi2009thermoelectric}
Dubi,~Y.; Di~Ventra,~M. Thermoelectric effects in nanoscale junctions. \emph{Nano letters} \textbf{2009}, \emph{9}, 97--101\relax
\mciteBstWouldAddEndPuncttrue
\mciteSetBstMidEndSepPunct{\mcitedefaultmidpunct}
{\mcitedefaultendpunct}{\mcitedefaultseppunct}\relax
\EndOfBibitem
\bibitem[Dubi and Di~Ventra(2011)Dubi, and Di~Ventra]{dubi2011colloquium}
Dubi,~Y.; Di~Ventra,~M. Colloquium: Heat flow and thermoelectricity in atomic and molecular junctions. \emph{Reviews of Modern Physics} \textbf{2011}, \emph{83}, 131\relax
\mciteBstWouldAddEndPuncttrue
\mciteSetBstMidEndSepPunct{\mcitedefaultmidpunct}
{\mcitedefaultendpunct}{\mcitedefaultseppunct}\relax
\EndOfBibitem
\bibitem[Ajisaka \latin{et~al.}(2015)Ajisaka, {\v{Z}}unkovi{\v{c}}, and Dubi]{ajisaka2015molecular}
Ajisaka,~S.; {\v{Z}}unkovi{\v{c}},~B.; Dubi,~Y. The molecular photo-cell: Quantum transport and energy conversion at strong non-equilibrium. \emph{Scientific reports} \textbf{2015}, \emph{5}, 8312\relax
\mciteBstWouldAddEndPuncttrue
\mciteSetBstMidEndSepPunct{\mcitedefaultmidpunct}
{\mcitedefaultendpunct}{\mcitedefaultseppunct}\relax
\EndOfBibitem
\bibitem[Sarkar and Dubi(2020)Sarkar, and Dubi]{sarkar2020environment}
Sarkar,~S.; Dubi,~Y. Environment-assisted and environment-hampered efficiency at maximum power in a molecular photocell. \emph{The Journal of Physical Chemistry C} \textbf{2020}, \emph{124}, 15115--15122\relax
\mciteBstWouldAddEndPuncttrue
\mciteSetBstMidEndSepPunct{\mcitedefaultmidpunct}
{\mcitedefaultendpunct}{\mcitedefaultseppunct}\relax
\EndOfBibitem
\bibitem[Nicolay \latin{et~al.}(2001)Nicolay, Reinert, H{\"u}fner, and Blaha]{nicolay2001spin}
Nicolay,~G.; Reinert,~F.; H{\"u}fner,~S.; Blaha,~P. Spin-orbit splitting of the L-gap surface state on Au (111) and Ag (111). \emph{Physical Review B} \textbf{2001}, \emph{65}, 033407\relax
\mciteBstWouldAddEndPuncttrue
\mciteSetBstMidEndSepPunct{\mcitedefaultmidpunct}
{\mcitedefaultendpunct}{\mcitedefaultseppunct}\relax
\EndOfBibitem
\bibitem[Zerah-Harush and Dubi(2018)Zerah-Harush, and Dubi]{zerah2018universal}
Zerah-Harush,~E.; Dubi,~Y. Universal origin for environment-assisted quantum transport in exciton transfer networks. \emph{The journal of physical chemistry letters} \textbf{2018}, \emph{9}, 1689--1695\relax
\mciteBstWouldAddEndPuncttrue
\mciteSetBstMidEndSepPunct{\mcitedefaultmidpunct}
{\mcitedefaultendpunct}{\mcitedefaultseppunct}\relax
\EndOfBibitem
\bibitem[Nitzan(2001)]{nitzan2001electron}
Nitzan,~A. Electron transmission through molecules and molecular interfaces. \emph{Annual review of physical chemistry} \textbf{2001}, \emph{52}, 681--750\relax
\mciteBstWouldAddEndPuncttrue
\mciteSetBstMidEndSepPunct{\mcitedefaultmidpunct}
{\mcitedefaultendpunct}{\mcitedefaultseppunct}\relax
\EndOfBibitem
\bibitem[Zhu(2004)]{zhu2004electronic}
Zhu,~X.-Y. Electronic structure and electron dynamics at molecule--metal interfaces: implications for molecule-based electronics. \emph{Surface Science Reports} \textbf{2004}, \emph{56}, 1--83\relax
\mciteBstWouldAddEndPuncttrue
\mciteSetBstMidEndSepPunct{\mcitedefaultmidpunct}
{\mcitedefaultendpunct}{\mcitedefaultseppunct}\relax
\EndOfBibitem
\bibitem[Dziarmaga \latin{et~al.}(2012)Dziarmaga, Zurek, and Zwolak]{dziarmaga2012non}
Dziarmaga,~J.; Zurek,~W.~H.; Zwolak,~M. Non-local quantum superpositions of topological defects. \emph{Nature Physics} \textbf{2012}, \emph{8}, 49--53\relax
\mciteBstWouldAddEndPuncttrue
\mciteSetBstMidEndSepPunct{\mcitedefaultmidpunct}
{\mcitedefaultendpunct}{\mcitedefaultseppunct}\relax
\EndOfBibitem
\bibitem[Kiran \latin{et~al.}(2016)Kiran, Mathew, Cohen, Hern{\'a}ndez~Delgado, Lacour, and Naaman]{kiran2016helicenes}
Kiran,~V.; Mathew,~S.~P.; Cohen,~S.~R.; Hern{\'a}ndez~Delgado,~I.; Lacour,~J.; Naaman,~R. Helicenes—A new class of organic spin filter. \emph{Advanced Materials} \textbf{2016}, \emph{28}, 1957--1962\relax
\mciteBstWouldAddEndPuncttrue
\mciteSetBstMidEndSepPunct{\mcitedefaultmidpunct}
{\mcitedefaultendpunct}{\mcitedefaultseppunct}\relax
\EndOfBibitem
\bibitem[Möllers \latin{et~al.}(2022)Möllers, Wei, Salamon, Bartsch, Wende, Waldeck, and Zacharias]{mollers2022spin2}
Möllers,~P.~V.; Wei,~J.; Salamon,~S.; Bartsch,~M.; Wende,~H.; Waldeck,~D.~H.; Zacharias,~H. Spin-polarized photoemission from chiral CuO catalyst thin films. \emph{ACS nano} \textbf{2022}, \emph{16}, 12145--12155\relax
\mciteBstWouldAddEndPuncttrue
\mciteSetBstMidEndSepPunct{\mcitedefaultmidpunct}
{\mcitedefaultendpunct}{\mcitedefaultseppunct}\relax
\EndOfBibitem
\bibitem[Monti and Steele(2010)Monti, and Steele]{monti2010influence}
Monti,~O.~L.; Steele,~M.~P. Influence of electrostatic fields on molecular electronic structure: insights for interfacial charge transfer. \emph{Physical Chemistry Chemical Physics} \textbf{2010}, \emph{12}, 12390--12400\relax
\mciteBstWouldAddEndPuncttrue
\mciteSetBstMidEndSepPunct{\mcitedefaultmidpunct}
{\mcitedefaultendpunct}{\mcitedefaultseppunct}\relax
\EndOfBibitem
\bibitem[Ernst(2023)]{baljozovic2023spin}
Ernst,~K.-H. Spin-and angle-resolved photoemission spectroscopy study of heptahelicene layers on Cu (111) surfaces. \emph{The Journal of Chemical Physics} \textbf{2023}, \emph{159}\relax
\mciteBstWouldAddEndPuncttrue
\mciteSetBstMidEndSepPunct{\mcitedefaultmidpunct}
{\mcitedefaultendpunct}{\mcitedefaultseppunct}\relax
\EndOfBibitem
\bibitem[Kondou \latin{et~al.}(2022)Kondou, Shiga, Sakamoto, Inuzuka, Nihonyanagi, Araoka, Kobayashi, Miwa, Miyajima, and Otani]{kondou2022chirality}
Kondou,~K.; Shiga,~M.; Sakamoto,~S.; Inuzuka,~H.; Nihonyanagi,~A.; Araoka,~F.; Kobayashi,~M.; Miwa,~S.; Miyajima,~D.; Otani,~Y. Chirality-Induced Magnetoresistance Due to Thermally Driven Spin Polarization. \emph{Journal of the American Chemical Society} \textbf{2022}, \emph{144}, 7302--7307\relax
\mciteBstWouldAddEndPuncttrue
\mciteSetBstMidEndSepPunct{\mcitedefaultmidpunct}
{\mcitedefaultendpunct}{\mcitedefaultseppunct}\relax
\EndOfBibitem
\bibitem[Trudel(2011)]{trudel2011unexpected}
Trudel,~S. Unexpected magnetism in gold nanostructures: making gold even more attractive. \emph{Gold Bulletin} \textbf{2011}, \emph{44}, 3--13\relax
\mciteBstWouldAddEndPuncttrue
\mciteSetBstMidEndSepPunct{\mcitedefaultmidpunct}
{\mcitedefaultendpunct}{\mcitedefaultseppunct}\relax
\EndOfBibitem
\bibitem[Li \latin{et~al.}(2011)Li, Wu, Karna, Wang, Hsu, Wang, and Li]{li2011intrinsic}
Li,~C.-Y.; Wu,~C.-M.; Karna,~S.~K.; Wang,~C.-W.; Hsu,~D.; Wang,~C.-J.; Li,~W.-H. Intrinsic magnetic moments of gold nanoparticles. \emph{Physical Review B} \textbf{2011}, \emph{83}, 174446\relax
\mciteBstWouldAddEndPuncttrue
\mciteSetBstMidEndSepPunct{\mcitedefaultmidpunct}
{\mcitedefaultendpunct}{\mcitedefaultseppunct}\relax
\EndOfBibitem
\bibitem[Sato \latin{et~al.}(2015)Sato, Ishikawa, Sato, and Sato]{sato2015magnetic}
Sato,~R.; Ishikawa,~S.; Sato,~H.; Sato,~T. Magnetic order of Au nanoparticle with clean surface. \emph{Journal of Magnetism and Magnetic Materials} \textbf{2015}, \emph{393}, 209--212\relax
\mciteBstWouldAddEndPuncttrue
\mciteSetBstMidEndSepPunct{\mcitedefaultmidpunct}
{\mcitedefaultendpunct}{\mcitedefaultseppunct}\relax
\EndOfBibitem
\bibitem[Nealon \latin{et~al.}(2012)Nealon, Donnio, Greget, Kappler, Terazzi, and Gallani]{nealon2012magnetism}
Nealon,~G.~L.; Donnio,~B.; Greget,~R.; Kappler,~J.-P.; Terazzi,~E.; Gallani,~J.-L. Magnetism in gold nanoparticles. \emph{Nanoscale} \textbf{2012}, \emph{4}, 5244--5258\relax
\mciteBstWouldAddEndPuncttrue
\mciteSetBstMidEndSepPunct{\mcitedefaultmidpunct}
{\mcitedefaultendpunct}{\mcitedefaultseppunct}\relax
\EndOfBibitem
\bibitem[Agrachev \latin{et~al.}(2017)Agrachev, Antonello, Dainese, Ruzzi, Zoleo, Aprà, Govind, Fortunelli, Sementa, and Maran]{agrachev2017magnetic}
Agrachev,~M.; Antonello,~S.; Dainese,~T.; Ruzzi,~M.; Zoleo,~A.; Aprà,~E.; Govind,~N.; Fortunelli,~A.; Sementa,~L.; Maran,~F. Magnetic ordering in gold nanoclusters. \emph{ACS omega} \textbf{2017}, \emph{2}, 2607--2617\relax
\mciteBstWouldAddEndPuncttrue
\mciteSetBstMidEndSepPunct{\mcitedefaultmidpunct}
{\mcitedefaultendpunct}{\mcitedefaultseppunct}\relax
\EndOfBibitem
\bibitem[Henk \latin{et~al.}(2003)Henk, Ernst, and Bruno]{henk2003spin}
Henk,~J.; Ernst,~A.; Bruno,~P. Spin polarization of the L-gap surface states on Au (111). \emph{Physical Review B} \textbf{2003}, \emph{68}, 165416\relax
\mciteBstWouldAddEndPuncttrue
\mciteSetBstMidEndSepPunct{\mcitedefaultmidpunct}
{\mcitedefaultendpunct}{\mcitedefaultseppunct}\relax
\EndOfBibitem
\end{mcitethebibliography}

\end{document}